\PassOptionsToPackage{table}{xcolor}
\documentclass[sigconf]{acmart}
\AtBeginDocument{%
  \providecommand\BibTeX{{%
    \normalfont B\kern-0.5em{\scshape i\kern-0.25em b}\kern-0.8em\TeX}}}

\usepackage{amsthm,amsmath,bm}
\usepackage{multirow}
\usepackage{mathrsfs}
\usepackage{graphicx}
\usepackage{subfigure} 
\usepackage{float}
\usepackage[export]{adjustbox} 
\usepackage{xspace}
\usepackage{adjustbox}
\graphicspath{ {./Figs/} }

\usepackage{algorithm}
\usepackage{algorithmic}
\usepackage{balance}
\usepackage{enumitem}

\newcommand{\eat}[1]{}
\newcommand{\paratitle}[1]{\vspace{1.5ex}\noindent\textbf{#1}}

\newcommand{\baby}{ADS\xspace}

\newcommand{\ignore}[1]{}

\copyrightyear{2025}
\acmYear{2025}
\setcopyright{acmlicensed}\acmConference[SIGIR '25]{Proceedings of the 48th International ACM SIGIR Conference on Research and Development in Information Retrieval}{July 13--18, 2025}{Padua, Italy}
\acmBooktitle{Proceedings of the 48th International ACM SIGIR Conference on Research and Development in Information Retrieval (SIGIR '25), July 13--18, 2025, Padua, Italy}
\acmDOI{XXXXXXX.XXXXXXX}
\acmISBN{978-1-4503-XXXX-X/18/06}

\begin{document}

\title{Adaptive Domain Scaling for Personalized Sequential \\Modeling in Recommenders}

\author{Zheng Chai}
\email{chaizheng.cz@bytedance.com}
\affiliation{%
  \institution{ByteDance}
  \city{Hangzhou}
  \country{China}
}

\author{Hui Lu}
\email{luhui.xx@bytedance.com}
\affiliation{%
  \institution{ByteDance}
  \city{Hangzhou}
  \country{China}
}

\author{Di Chen}
\email{chendi.666@bytedance.com}
\affiliation{%
  \institution{ByteDance}
  \city{Beijing}
  \country{China}
}

\author{Qin Ren}
\email{renqin.97@bytedance.com}
\affiliation{%
  \institution{ByteDance}
  \city{Beijing}
  \country{China}
}

\author{Yuchao Zheng\textsuperscript{†}}
\thanks{\textsuperscript{†}Corresponding Author.}
\email{zhengyuchao.yc@bytedance.com}
\affiliation{%
  \institution{ByteDance}
  \city{Beijing}
  \country{China}
}

\author{Xun Zhou}
\email{zhouxun@bytedance.com}
\affiliation{%
  \institution{ByteDance}
  \city{Beijing}
  \country{China}
}

\begin{abstract}
  Users generally exhibit complex behavioral patterns and diverse intentions in multiple business scenarios of super applications like Douyin, presenting great challenges to current industrial multi-domain recommenders. To mitigate the discrepancies across diverse domains, researches and industrial practices generally emphasize sophisticated network structures to accomodate diverse data distributions, while neglecting the inherent understanding of user behavioral sequence from the multi-domain perspective. In this paper, we present Adaptive Domain Scaling (\baby) model, which comprehensively enhances the personalization capability in target-aware sequence modeling across multiple domains. Specifically, \baby comprises of two major modules, including personalized sequence representation generation (PSRG) and personalized candidate representation generation (PCRG). The modules contribute to the tailored multi-domain learning by dynamically learning both the user behavioral sequence item representation and the candidate target item representation under different domains, facilitating adaptive user intention understanding. Experiments are performed on both a public dataset and two billion-scaled industrial datasets, and the extensive results verify the high effectiveness and compatibility of \baby. Besides, we conduct online experiments on two influential business scenarios including Douyin Advertisement Platform and Douyin E-commerce Service Platform, both of which show substantial business improvements. Currently, \baby has been fully deployed in many recommendation services at ByteDance, serving billions of users.
  
\end{abstract}

\ccsdesc[500]{Information systems~Recommender systems}

\keywords{Multi-Domain Learning, Sequential Modeling, Ranking, Personalized Recommender System}




\maketitle

\section{Introduction}

\begin{figure}[H]
\centering
\begin{minipage}[b]{\linewidth}
\subfigure[Short-Video]{
\includegraphics[width=0.31\linewidth,trim=90 0 90 0,clip]{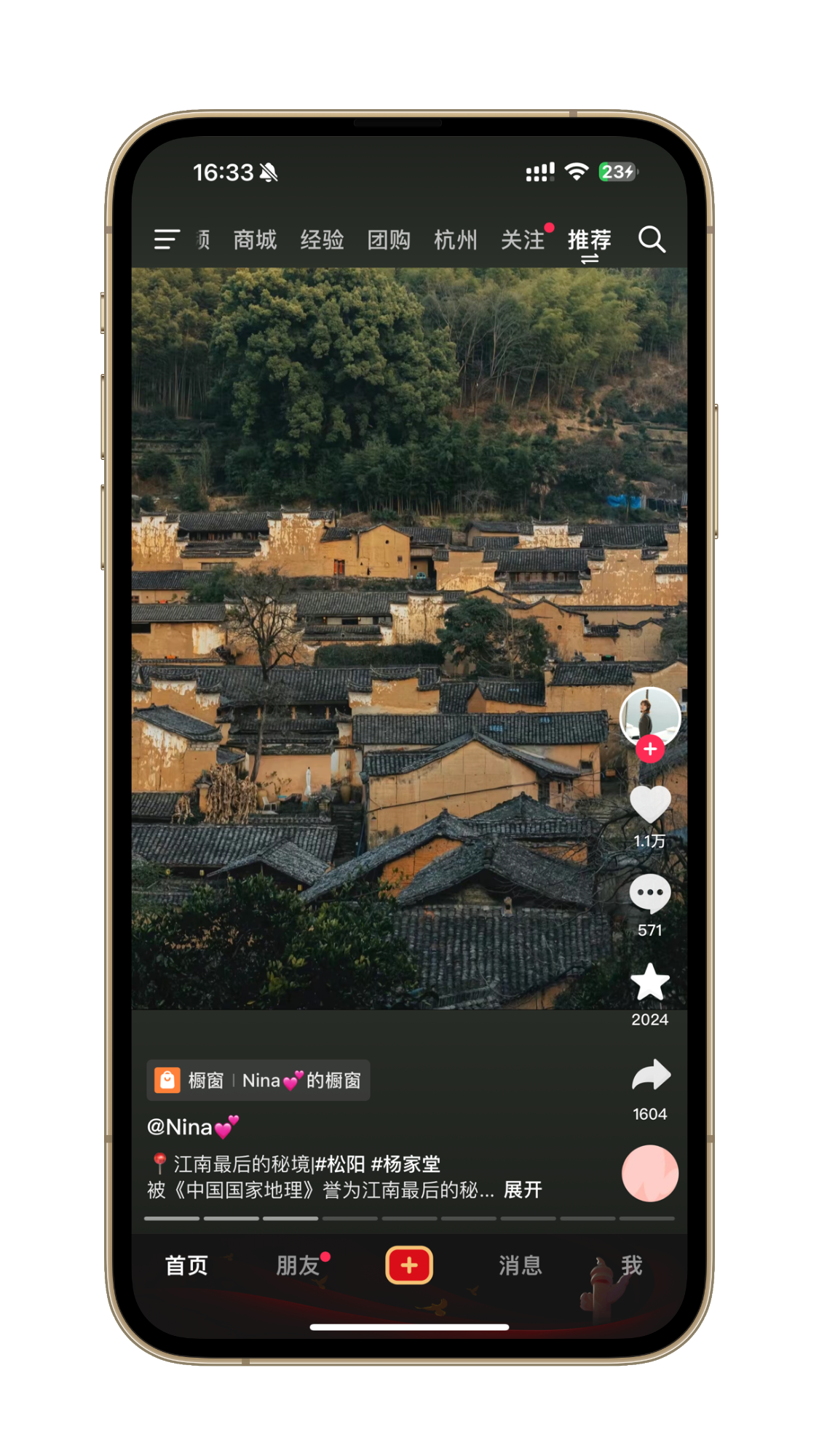}
}
\subfigure[Live-Preview]{
\includegraphics[width=0.31\linewidth,trim=90 0 90 0,clip]{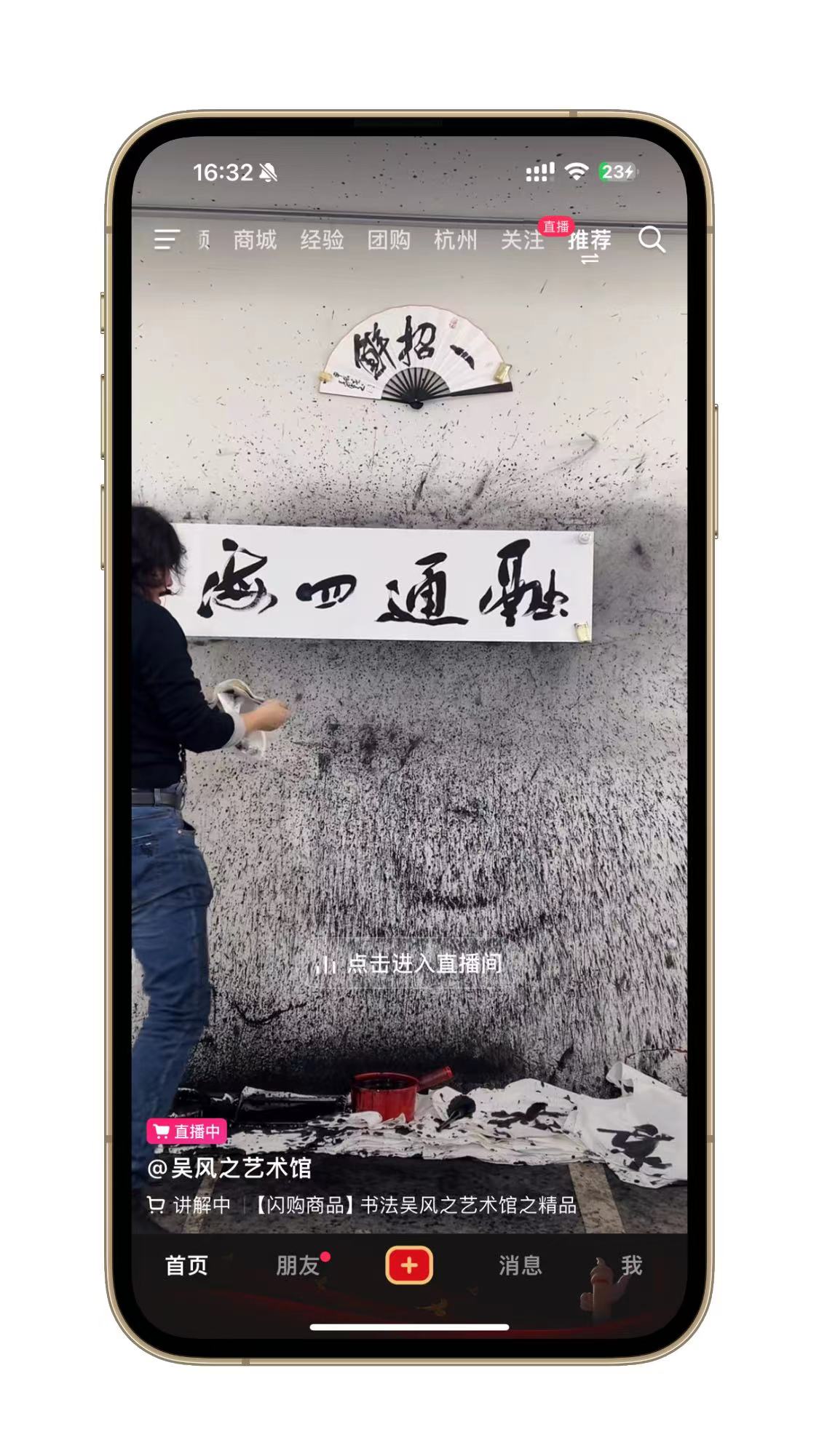}
}
\subfigure[Live-Slide]{
\includegraphics[width=0.31\linewidth,trim=90 0 90 0,clip]{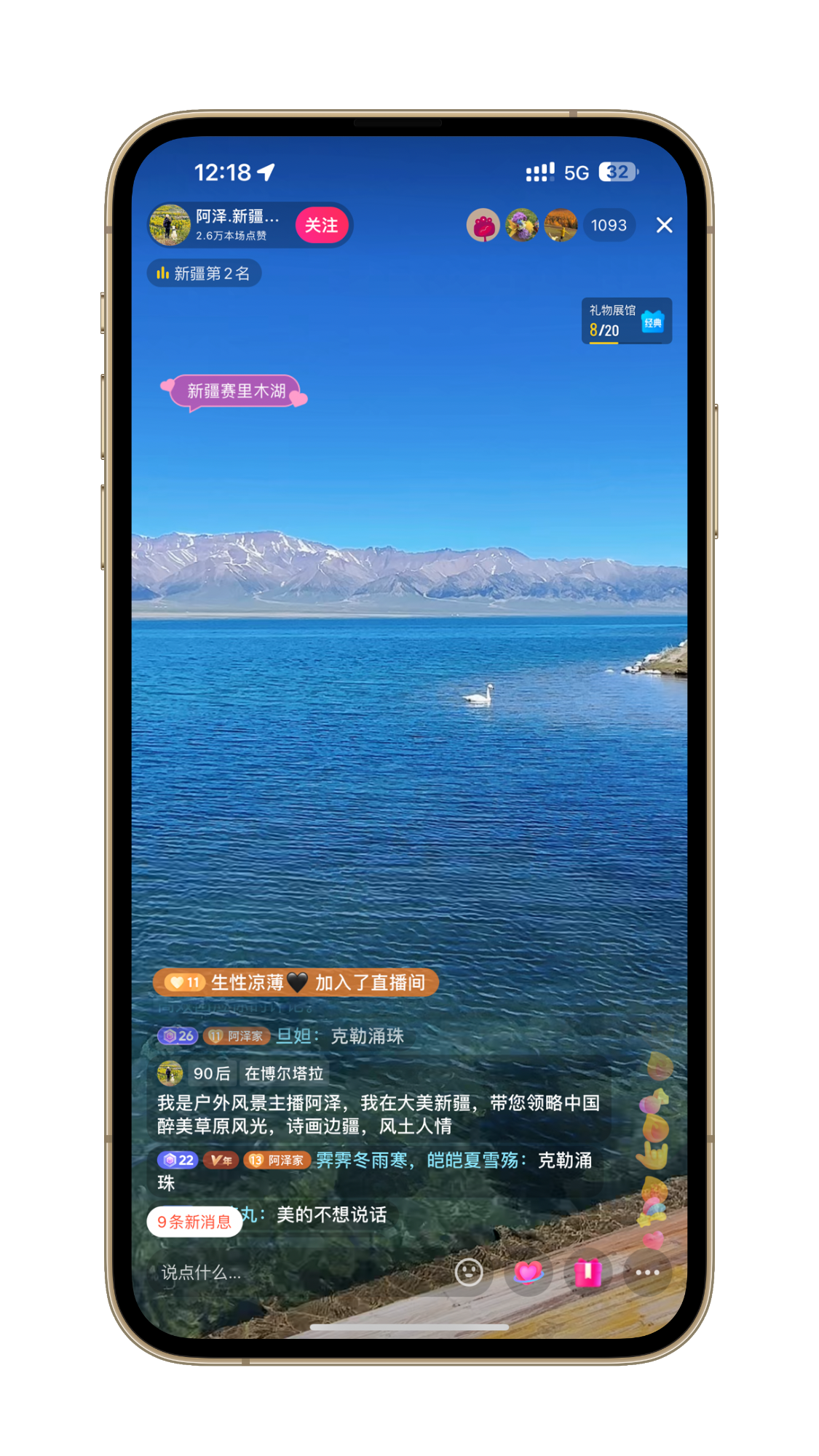}
}
\end{minipage}
\caption{Typical business scenarios in Douyin.}
\label{fig:scenarios}
\end{figure}

With the exponential growth of digital contents and the widespread use of the internet, recommender systems have become a vital role for enhancing user experience and alleviating information overload \cite{zhang2019deep, chai2022umi, lu2025large}. In real-world applications, for improving user retention and promote business benefit, the demands of industrial recommendation are widely distributed cross multiple  domains \cite{jiang2022adaptive, li2023one}. For example, as shown in Fig~\ref{fig:scenarios}, in Douyin\footnote{https://www.douyin.com/}, one of the largest video-watching apps in the world, the major domains include \textit{Short-videos, Live-preview}, and \textit{Live-slide}, where users can watch short-videos, live-streams, and enjoy the e-commerce and local-life services. Besides, due to its billion-scale user volume, different user groups, like users from different countries, with different genders, highly-active or not, also contribute to different domains. As the data distrbutions are quite diversed across different domains, it poses a significant multi-domain modeling problem for the recommender system \cite{zhang2024multi}.

To this end, \textit{de facto} common industrial practices generally build a shared-bottom with multi-heads outputs model structure, leveraging the advantages of both separate and unified mixup modeling for multiple domains \cite{caruana1993multitask}. To further improve this, recent approaches have put efforts in building elaborate network structures to enhance multi-domain modeling, e.g., the domain-level methods like the star topology adaptive recommender (STAR) \cite{sheng2021one}, progressive layered extraction (PLE) \cite{tang2020progressive}, and the instance-level methods like adaptive parameter generation network (APG) \cite{yan2022apg}, AdaSparse \cite{yang2022adasparse}, etc. However, most of the existing approaches are designed for sophisticated feature interaction network structures, while the approaches to multi-domain sequential modeling attract much less attention.



Sequential modeling plays a vital role in industrial recommenders, among which the most popular and effective methods are the target-aware attention based methods, for example, deep interest network (DIN) \cite{zhou2018deep}, feature co-action network (CAN) \cite{bian2022can}, and multi-head attention (MHA) \cite{vaswani2017attention}. Despite its significance, the impacts of multi-domain discrepancies are less considered in existing target attention methods, leaving a remarkable gap for the area of multi-domain modeling. Generally speaking, the current target attention mechanism for user sequence can be formulated as a typical query-key-value modeling paradigm: $g(Rep_{cand}, Rep_{seq}) \times Rep_{seq}$, where $Rep_{cand}$ denotes the representation of the candidate target item whose click/convert probabilities need to be predicted, $Rep_{seq}$ indicates the user sequence embeddings, and $g$ calculates the attention weight between any sequence-item and target-item pairs. As discussed previously, current industrial recommenders generally follow a shared-bottom embedding paradigm, which means that 1) the embedding table of both the candidate item and user behavioral items are fully shared, with no consideration of the distinctions between items and users belonging to different domains, and 2) the candidate item serves as a shared query for different keys/values, with no consideration of the distinctions between the multiple items in user sequence which generally occurs across multiple domains. Accordingly, this poses potential challenges to the current multi-domain recommenders from two aspects:


\begin{itemize}[leftmargin=*]
    \item \textbf{Personalization of the Sequence Representations}. The multi-domain representations of identical items occurred in different users' sequences are of necessity in recommenders. For example, new users would like to watch the highly-liked videos, while some long-time users may pay attention to the video creators they follow. Thus, the same video shows different attractions to various user domains, while its embedding is a shared representation among different user sequences, which hinders the recommender to grasp the actual intention of the user.
    \item \textbf{Personalization of the Candidate Item}. For different users or different items in the same user's sequence, the candidate item has different influence and function due to the multi-domain impact. For example, user's shopping behaviors in Douyin Mall might be primarily influenced by product prices, while the video creators have a more significant impact for content-preferred users in Douyin short-videos scenario. Thus an identical candidate item should be personalized across domains to accommodate different historical items in user sequence.
\end{itemize}

To overcome the limitations and fulfill the gap in multi-domain target-aware attention modeling, we propose Adaptive Domain Scaling (\baby) model, which fully mines the personalization modeling ability of the current target-attention based recommenders and provides more accurate and adaptive intention understanding ability in multi-domain tasks. Specifically, \baby comprises two modules, namely, Personalized Sequence Representation Generation (PSRG) and Personalized Candidate Representation Generation (PCRG). In PSRG, we designed a novel shared-and-private structure for learning multi-domain item representations in user behaviors, which aims to generate personalized representations for sequence items, i.e., the same item occured in different user's sequence has different representations. In PCRG, the candidate item further enhances the personalization modeling ability via generating different target candidate representations for different sequence items. With the domain-related information as input for the generation structures, the impact of multi-domains are sufficiently injected to the sequence modeling, and thus enhances user intention understanding. Note that \baby is an efficient plug-and-play network, and can be readily integrated into current recommenders.

The contribution of this work can be summarized as follows:
\begin{itemize}[leftmargin=*]
\item We present Adaptive Domain Scaling (ADS) model, an effective plug-and-play personalization network structure for multi-domain user intent understanding by personalizing the target-aware attention modeling. We conduct extensive experiments on both a public dataset and two billion-scaled industrial dataset, and the results verify its superiority.
\item Both personalized sequence representation generation and personalized candidate representation generation modules are developed in our framework, which captures the multi-domain characteristics from the viewpoint of users behavioral sequences and candidate target item, enhancing the multi-domain learning efficacy for current target-aware attention mechanisms.

\item We deploy \baby in both the advertisement system and the e-commercial system of Douyin at ByteDance, which brings significant 1.00\% and 0.79\% lifts of total revenue in the Douyin ads system and the e-commercial system, respectively. Currently, \baby has been fully deployed in many recommendation systems at ByteDance, serving billions of users.
\end{itemize}

\section{Methodology}
\subsection{Preliminaries}
\subsubsection{Problem Formulation}
In this paper, we focus on the ranking modeling task in recommenders, which is a typical binary classification problem. Generally, taking the prediction of the click-through rate (CTR) as an example, the probability $\hat{y}$ can be obtained via the following:

\begin{figure*}[h]
\center
\includegraphics[width=\linewidth]{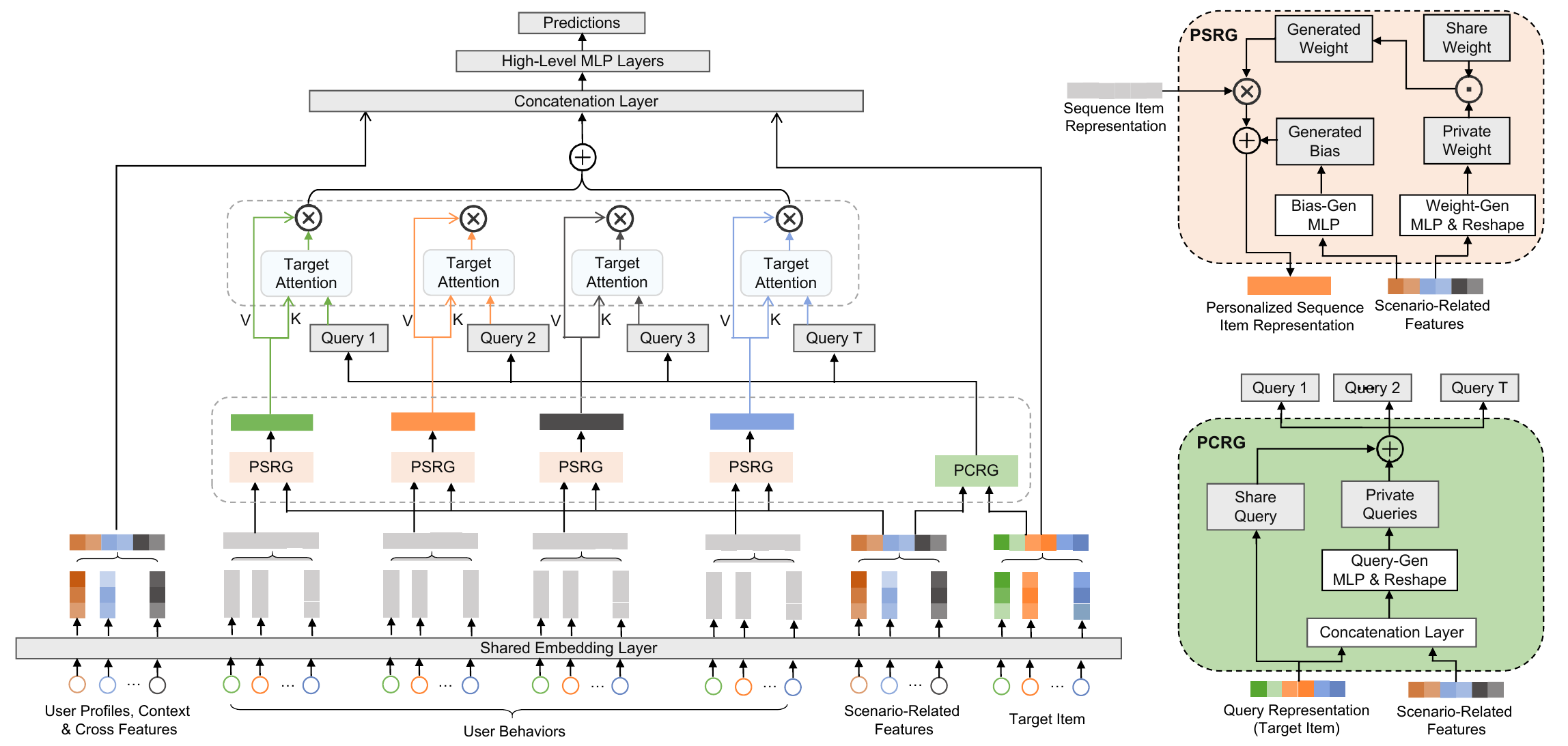}
\caption{Overview of \baby. \baby consists of PCRG, PSRG, and target attention module. Given scenario-related features and target item as input, PCRG first generates multiple queries considering the co-pattern of the target item (query) and the scenario features. For PSRG, it takes scenario features as input to generate weight and bias parameters to formulate the personalized MLP, then the original sequence item embedding is passed through the generated MLP to obtain personalized representation. Both the PCRG and PSRG share a share-and-private learning paradigm. Finally, the generated sequence is aggregated by the generated multi-queries with the target-aware attention mechanism, and the concatenation layer and high-level MLP layers are finally used to make predictions.}
\label{fig:overview}
\end{figure*}

\begin{equation}
    \hat{y} = f(\bm{E}_U, \bm{E}_I, \bm{E}_O)
\end{equation}

\noindent in which $\bm{E}(\cdot)$ indicates the embedding function, in which the raw categorical features are translated directly into embeddings, and the continuous features are first bucketed and then embedded as dense vectors. $f$ is the MLP-based transformation function. $U$, $I$, and $O$ denote the user-side, target candidate item-side, and other features, respectively. User-side features generally consist of demographic feature set (for example, user locations and languages) and behavioral feature set (for example, the watch list or the shopping list of users). Item-side features include the item's descriptive features like its category, creators, etc. Besides, the other features $O$ generally contains contextual and user-item cross features.

\subsection{The Proposed ADS}
The structure of the proposed ADS is illustrated in Figure~\ref{fig:overview}. Overall, it is composed of two major parts:
\begin{itemize}[leftmargin=*]
    \item \textbf{Personalized Sequence Representation Generation}. PSRG generates dynamic behavioral item embedding with a share-and-private learning structure, such that the same item occured in different domains has unshared representations due to the distinction among multi-domains.

    \item \textbf{Personalized Candidate Representation Generation}. PCRG captures different aspects of sequence items and generates multiple adaptive queries (i.e., candidate items) for each sequence item, such that the different query influence on diverse sequence items can be reflected.
\end{itemize}

With the adaptive queries, keys, and values generated by PCRG and PSRG, the target-aware modeling mechanism like MHA, DIN, and CAN, can be readily integrated into this framework, facilitating the interest capture in multi-domain scenarios.

\subsubsection{Personalized Sequence Representation Generation (PSRG)}
Current large-scale industrial recommenders generally adopt the share-embedding layer to embed the raw ID and other features into dense vectors. In this manner, a specific item in the embedding table has a unified embedding, which is shared across different user sequences and neglects the impact from the difference between multiple domains.

The basic idea of PSRG is to dynamically generate a personalized layer for each item embedding in user behavioral sequence, such that the original shared representation can be diversified across multiple domains. Specifically, we use the concatenation of the domain-related features embeddings $\bm{E}_D \in \mathbb{R}^{d_D}$ as the input of the generation part of PSRG, which consists of the following two feature categories: 1) explicit-domain-indicator features, which distinguish the domain that a sample belongs to. For example, the indicator ranges from $[0, 2]$ to indicate the three distinct business scenarios within Douyin; and 2) implicit-domain-indicator features. In recommender systems, some domains can be challenging to define explicitly. For example, whether a user is highly-active or not. Therefore, additional engineer-constructed statistical features are necessary to be incorporated to further capture and differentiate various domains. With the two categories of features, as illustrated in Figure~\ref{fig:overview}, the generation processes of the weight and bias for sequence items are designed to dynamically adjust the original item embedding.

\textbf{Sequence-Weight Gen-Net}. Denote the user sequence embedding as $\bm{E}_S \in \mathbb{R}^{T \times d_S}$, where $T$ and $d_S$ represent the user sequence length and the embedding dimension of each sequence item, respectively. Based on the domain features $\bm{E}_D$, the weight generation process consists of a \textit{private} weight part and a \textit{share} weight part to capture both the commonality and individuality of multi domains. For the private part, a two-layered MLP is performed to generate the private weight:

\begin{align}
\mathbf{W}_{private} = Sigmoid({\rm ReLU}(\bm{E}_D \mathbf{W}_1^{\rm{T}} + \bm{b}_1)\mathbf{W}_2^{\rm{T}} + \bm{b}_2)
\label{eqn:private_weight}
\end{align}

\noindent where $\mathbf{W}_1^{\rm{T}} \in \mathbb{R}^{d_D \times d_h}$, $\mathbf{W}_2^{\rm{T}} \in \mathbb{R}^{d_h \times (d_S \times d_S)}$, $\bm{b}_1 \in \mathbb{R}^{d_h}$, $\bm{b}_2 \in \mathbb{R}^{(d_S \times d_S)}$, and $d_h$ denotes the hidden layer dimension. Note that the two-layer function instead of a single layer can not only improve the expression ability of the model, but also significantly reduce the model parameters and computational costs, as $d_S$ is generally at the order of tens in practical cases, and $d_h << (d_S \times d_S)$. 

Based on $\mathbf{W}_{private}$, a global weight $\mathbf{W}_{shared} \in \mathbb{R}^{(d_S \times d_S)}$ is further defined as a learnable matrix which is shared across all the users. To enjoy the merits in learning both the commonality and individuality, the generated weight is defined as:

\begin{align}
\mathbf{W}_{generated} = \eta * (\mathbf{W}_{shared}  \odot \mathbf{W}_{private})
\label{eqn:generated_weight}
\end{align}

\noindent where $\odot$ denotes the element-wise product. As the value of $\mathbf{W}_{private}$ ranges from $[0, 1]$ due to introduction of $Sigmoid$, a scaling hyperparameter $\eta$ is further involved to enlarge the expression range of $\mathbf{W}_{private}$. Therefore, the calculation in Equation~\ref{eqn:generated_weight} can be viewed as a adaptive scaling method for the global parameter $\mathbf{W}_{shared}$.

\textbf{Sequence-Bias Gen-Net}.
Similar to the above weight generation process, the bias generation can be readily obtained via the following equation:

\begin{align}
\bm{b}_{generated} = {\rm ReLU}(\bm{E}_D {\mathbf{W}_{1}^{\prime}}^{\rm{T}} + \bm{b}_{1}^{\prime}) {\mathbf{W}_{2}^{\prime}}^{\rm{T}} + \bm{b}_{2}^{\prime}
\label{eqn:generate_bias}
\end{align}

\noindent where ${\mathbf{W}_{1}^{\prime}}^{\rm{T}} \in \mathbb{R}^{d_D \times d_h}$, ${\mathbf{W}_{2}^{\prime}}^{\rm{T}} \in \mathbb{R}^{d_h \times d_S}$, $\bm{b}_1^{\prime} \in \mathbb{R}^{d_h}$, and $\bm{b}_2^{\prime} \in \mathbb{R}^{d_S}$. With the generated weight and bias, the PSRG can be achieved via the following:

\begin{align}
\bm{E}_{S-personalized} = \bm{E}_{S} {{Reshape}} (\mathbf{W}_{generated})^{\rm{T}} + \bm{b}_{generated}
\label{eqn:psrg}
\end{align}

\noindent where the ${Reshape}$ operator refers to reshape the 1-D vector-form $\mathbf{W}_{generated}$ as a 2-D matrix-form with the shape of $d_S \times d_S$.

\subsubsection{Personalized Candidate Representation Generation (PCRG)}
In addition to the personalized modeling for sequence, the other important part is the multi-domain modeling for the target item, which generally plays the role of query in target-aware attention. Typically, personalizing the candidate item encompasses two aspects. On one hand, similar to the sequence representation, the representation of the target item itself is also embedded through the shared embedding layer, which is not personalized across different domains. On the other hand, the candidate item plays different roles for different sequence items in various domains. For example, user's watchlist in Douyin mall channel reflects her shopping interests, while in short-video channel reflects her content preference. Thus, user's diverse interest should be captured via a more personalized query.


\begin{figure}
\center
\includegraphics[width=\linewidth, trim=48 343 520 43, clip]{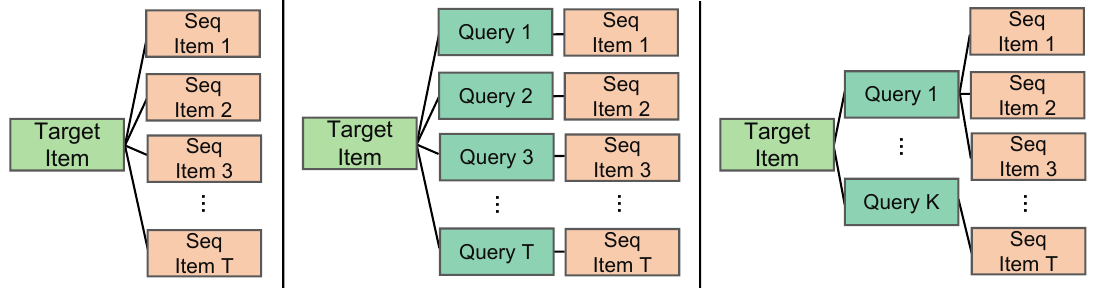}
\caption{Comparison of traditional target-attention methods (Left) and the Multi-Query Gen-Net (Middle and Right).}
\label{fig:multi_query}
\end{figure}
\textbf{Multi-Query Gen-Net}. To this end, we propose \textit{Multi-Query Gen-Net}, as shown in the middle subfigure in Figure~\ref{fig:multi_query}, which produces multiple queries corresponding to different sequence items, under the guidance of the domain-related features $\bm{E}_D$ and the original target item embedding $\bm{E}_Q \in \mathbb{R}^{d_Q}$:

\begin{align}
\bm{E}_{Q-private} = {\rm{ReLU}} \left( \left(\bm{E}_D \oplus \bm{E}_Q \right) \mathbf{W}_{q1}^{\rm{T}} + \bm{b}_{q1} \right) \mathbf{W}_{q2}^{\rm{T}} + \bm{b}_{q2}
\label{eqn:q_private}
\end{align}

\noindent where $\oplus$ refers to the concatenation operation,  $\mathbf{W}_{q1}^{\rm{T}} \in \mathbb{R}^{(d_D+d_Q) \times d_h}$, $\bm{b}_{q1} \in \mathbb{R}^{d_h}$, $\mathbf{W}_{q2}^{\rm{T}} \in \mathbb{R}^{d_h \times (T \times d_Q)}$ and $ \bm{b}_{q2} \in \mathbb{R}^{T \times d_Q}$. Noted that the hidden layer dimension $d_h << (T \times d_Q)$, yielding a controllable computation cost at $\bm{\mathcal{O}}(d_h T d_Q)$. 


\textbf{Chunked-Query Generation}. For cases with long sequence with larger $T$ at hundreds or even higher, we also devise a lightweight chunked-query generation method for improved computation efficiency. As shown in the right subfigure in Figure~\ref{fig:multi_query}, since user's adjacent actions are prone to happen in the same domain, the raw sequence can be divided into $G$ chunks with the adjacent items formed as a group.\footnote{Without losing generality, here it is assumed that $T$ is divisible by $G$ with padding items.} Therefore, the generated $\bm{E}_{Q-private} \in \mathbb{R}^{(G \times d_Q)}$ can be further repeated to $\mathbb{R}^{(T \times d_Q)}$ and the cost further reduces to $\bm{\mathcal{O}}(d_h G d_Q)$.

Corresponding to the multiple private queries $\bm{E}_{Q-private} \in \mathbb{R}^{(T \times d_Q)}$, we use the original query $\bm{E}_Q$ as the shared base, i.e., $\bm{E}_{Q-shared} = tile(\bm{E}_Q)$, where $tile$ refers to the tile operator which repeat the $\bm{E}_Q$ by $T$ times, i.e., $\bm{E}_{Q-shared} \in \mathbb{R}^{(T \times d_Q)}$. Then, the final generated multiple queries can be obtained via a residual manner:

\begin{align}
\bm{E}_{Q-personalized} = Reshape(\bm{E}_{Q-private} + \bm{E}_{Q-shared})
\label{eqn:q_generated}
\end{align}

\noindent where $Reshape$ operator reshapes the 1-D vector as a 2-D matrix-from with the shape of $T \times d_Q$.

\subsubsection{Target-Aware Attention \& Prediction}
With the above personalized queries $\bm{E}_{Q-personalized} \in \mathbb{R}^{T \times d_Q}$ and personalized sequence items $\bm{E}_{S-personalized} \in \mathbb{R}^{T \times d_{S}}$, the target attention module is performed to calculate the attention weight of each item and aggregate the sequence under the guidance of queries. Generally, the personalized queries and items can be readily integrated into many popular attention methods, like multi-head target attention, DIN, and CAN. Taking the multi-head target attention as an example, for each head, the candidate item and sequence items are first transformed via the following:

\begin{align}
\bm{Q} = \bm{E}_{Q-personalized}W_Q \\
\bm{K} = \bm{E}_{S-personalized}W_K \\
\bm{V} = \bm{E}_{S-personalized}W_V
\label{eqn:qkv}
\end{align}

\noindent where $W_Q \in \mathbb{R}^{d_Q \times d_A}$, $W_K$ and $W_V \in \mathbb{R}^{d_S \times d_A}$, in which $d_A$ refers to the dimension size in target attention. The attention weight $\bm{z}^{\prime}[t]$ in the $t$-th query-key pair, i.e., $\{\bm{Q}_t \in \mathbb{R}^{d_A}, \bm{K}_t \in \mathbb{R}^{d_A}\}$, can then be obtained by:

\begin{align}
\bm{z}^{\prime}[t] = \frac{\bm{Q}_t ^{\textsc{T}} \bm{K}_t}{\sqrt{d_A}}, \quad \rm{where}\ 1 \leq t \leq T
\label{eqn:attention}
\end{align}

\noindent following which a softmax-based operation is performed to normalize the personalized weights and aggregate the personalized sequence:

\begin{align}
\bm{z} = softmax(z^{\prime}),   \quad \bm{s} = \Sigma_{t=1}^{T} (z[t] \cdot \bm{V}_t)
\label{eqn:weighted_sum}
\end{align}

With the sequence modeling output $\bm{s}$ and the other feature embeddings including $\bm{E}_U$, $\bm{E}_I$, and $\bm{E}_O$, a concatenation layer and several high-level MLPs are performed to merge all the information and output the prediction, and the training loss can be obtained via the binary cross-entropy function.

\begin{align}
\bm{E}_{all} = \bm{s} \oplus \bm{E}_U \oplus \bm{E}_I \oplus \bm{E}_O, \quad \hat{y} = MLP(\bm{E}_{all})
\label{eqn:concat}
\end{align}



\section{Experiments}

\begin{table}
\caption{Statistics of the two datasets.}
\begin{tabular}{lcccc}
\hline
Dataset & $\#$Users & $\#$Items & $\#$Instances & $\#$Tasks \\ \hline
Taobao & 101,342 & 500,272 & 24.02M & Order \\
Douyin-Ads & 688M & 270M & 2.52B & CVR \\
Douyin-Ecom & 596M & 12.91M & 24.65B & Click \& Order \\ \hline
\end{tabular}
\label{tab:stats}
\end{table}

\begin{table*}[]
\caption{Comparison results of different methods on the three datasets. "D1", "D2", and "D3" are the abbreviations of "Domain 1", "Domain 2", and "Domain 3". Boldface denotes the best results in each group, and the boldface in the gray shadow area denotes that \baby significantly outperforms the second-best approach at the level of ($p < $ 0.05) in each group. For the two industrial datasets, note that a \underline{0.1\% Overall Imp.} in Douyin Ads and \underline{0.2\% Overall Imp.} in Douyin Ecom is considered to be significant improvement that can affect the performance of online A/B tests.}
\begin{adjustbox}{max width=\textwidth}
\begin{tabular}{c|cc|ccccc|cccccccc}
\hline
& \multicolumn{2}{c|}{{Taobao}} & \multicolumn{5}{c|}{{Douyin Ads}}                                                                                                                                  & \multicolumn{8}{c}{{Douyin E-commerce}}                                                                                                                                                                                                                                                                                                                                 \\ \cline{2-16} 
 
{ }                         & \multicolumn{2}{c|}{{Order}}  
                                & \multicolumn{5}{c|}{{CVR}}                                                                                                                                         & \multicolumn{4}{c|}{{Click}}                                                                                                                                                       & \multicolumn{4}{c}{{ Order}}                                                                                                           \\ \cline{2-16} 
 
\multirow{-3}{*}{{ Models}} & { Overall} & { \begin{tabular}[c]{@{}c@{}}Overall \\ Imp.\end{tabular}} & { D1}     & { D2}     & { D3}     & { Overall} & { \begin{tabular}[c]{@{}c@{}}Overall \\ Imp.\end{tabular}} & { D1}     & { D2}     & { Overall} & \multicolumn{1}{c|}{{ \begin{tabular}[c]{@{}c@{}}Overall \\ Imp.\end{tabular}}} & { D1}     & { D2}     & { Overall} & { \begin{tabular}[c]{@{}c@{}}Overall \\ Imp.\end{tabular}} \\ \hline
{ DNN}                                              & {0.6490} & {-} & { 0.8224} & { 0.8221} & { 0.8717} & { 0.8461}  & { -}                                                       & { 0.7949} & { 0.8418} & { 0.9082}  & \multicolumn{1}{c|}{{ -}}                                                          & { 0.8456} & { 0.8494} & { 0.8439}  & { -}                                                          \\
 
{ DeepFM}                                           & {0.6503} & {+0.87\%} & { 0.8229} & { 0.8233} & { 0.8720} & { 0.8469}  & { +0.23\%}                                        & { 0.7953} & { 0.8422} & { 0.9083}  & \multicolumn{1}{c|}{{ +0.02\%}}                                                    & { 0.8458} & { 0.8496} & { 0.8442}  & { +0.08\%}                                                    \\
 
{ DCNv2}                                            & {0.6505} & {+1.01\%} & { \textbf{0.8232}} & {\textbf{0.8235}} & {\textbf{0.8725}} & { \textbf{0.8470}}  & {\textbf{+0.26\%}}                                        & { 0.7950} & { 0.8418} & { 0.9082}  & \multicolumn{1}{c|}{{ +0.00\%}}                                                    & { 0.8456} & { 0.8494} & { 0.8439}  & { +0.00\%}                                                    \\
 
{ APG}                                              & {0.6484} & {-0.40\%} & { 0.8225} & { 0.8223} & { 0.8720} & { 0.8464}  & { +0.08\%}                                        & { 0.7951} & { 0.8418} & { 0.9082}  & \multicolumn{1}{c|}{{ +0.00\%}}                                                    & { 0.8456} & { 0.8494} & { 0.8439}  & { +0.00\%}                                                    \\
 
{ AdaSparse}                                        & {\textbf{0.6498}} & {\textbf{+0.54\%}} &{ 0.8228} & { 0.8230} & { 0.8721} & { 0.8465}  & { +0.11\%}                                        & { 0.7951} & { 0.8418} & { 0.9082}  & \multicolumn{1}{c|}{{ +0.00\%}}                                                    & { 0.8455} & { 0.8491} & { 0.8438}  & { -0.03\%}                                                    \\
 
{ DFFM}                                             & {0.6487} & {-0.20\%} & { 0.8232} & { 0.8241} & { 0.8722} & { 0.8465}  & { +0.11\%}                                        & { \textbf{0.7961}} & { \textbf{0.8440}} & { \textbf{0.9089}}  & \multicolumn{1}{c|}{{ \textbf{+0.17\%}}}                                                    & { \textbf{0.8469}} & { \textbf{0.8507}} & { \textbf{0.8455}}  & { \textbf{+0.46\%}}         \\ \hline
 
\multicolumn{1}{l|}{{ DIN}}                                              & {0.6498} & {+0.54\%} & { 0.8235} & { 0.8251} & { 0.8724} & { 0.8469}  & { +0.23\%}                                        & { 0.7967} & { 0.8443} & { 0.9091}  & \multicolumn{1}{c|}{{ +0.22\%}}                                                    & { 0.8472} & { 0.8509} & { 0.8456}  & { +0.49\%}                                                    \\
 
{ +FRNet}                                        & {0.6490} & {+0.00\%} & { 0.8235} & { 0.8247} & { 0.8722} & { 0.8469}  & { +0.23\%}                                        & { 0.7967} & { 0.8444} & { 0.9092}  & \multicolumn{1}{c|}{{ +0.24\%}}                                                    & { 0.8474} & { 0.8511} & { 0.8458}  & { +0.55\%}                                                    \\
 
{ +PEPNet}                                       & {0.6500} & {+0.67\%} & { 0.8235} & { 0.8250} & { 0.8723} & { 0.8470}  & { +0.26\%}                                        & { 0.7964} & { 0.8440} & { 0.9090}  & \multicolumn{1}{c|}{{ +0.19\%}}                                                    & { 0.8472} & { 0.8507} & { 0.8456}  & { +0.49\%}                                                    \\
\rowcolor[HTML]{EFEFEF} 
{ +\baby}                                       & { \textbf{0.6507}} & {\textbf{+1.14\%}} & {\textbf{0.8245}} & { \textbf{0.8262}} & { \textbf{0.8733}} & { \textbf{0.8477}}  & { \textbf{+0.46\%}}                                        & { \textbf{0.7972}} & { \textbf{0.8451}} & { \textbf{0.9094}}  & \multicolumn{1}{c|}{\cellcolor[HTML]{EFEFEF}{ \textbf{+0.29\%}}}                                                    & { \textbf{0.8477}} & { \textbf{0.8513}} & { \textbf{0.8462}}  & { \textbf{+0.66\%}}                                                    \\ \hline
\multicolumn{1}{l|}{{ MHA}}                                              & {0.6498} & {+0.54\%} & { 0.8232} & { 0.8238} & { 0.8721} & { 0.8467}  & { \textbf{+0.17\%}}                                        & { 0.7964} & { 0.8439} & { 0.9090}  & \multicolumn{1}{c|}{{ +0.19\%}}                                                    & { 0.8465} & { 0.8503} & { 0.8450}  & { +0.31\%}                                                    \\
 
{ +FRNet}                                        & {0.6496} & {+0.40\%} & { 0.8233} & { 0.8240} & { 0.8722} & { 0.8467}  & { +0.17\%}                                        & { 0.7963} & { 0.8437} & { 0.9089}  & \multicolumn{1}{c|}{{ +0.17\%}}                                                    & { 0.8468} & { 0.8506} & { 0.8453}  & { +0.40\%}                                                    \\
 
{ +PEPNet}                                       & {0.6498} & {+0.54\%} & { 0.8237} & { 0.8246} & { 0.8725} & { 0.8471}  & { +0.28\%}                                        & { 0.7964} & { 0.8439} & { 0.9090}  & \multicolumn{1}{c|}{{ +0.19\%}}                                                    & { 0.8467} & { 0.8505} & { 0.8452}  & { +0.37\%}                                                    \\
\rowcolor[HTML]{EFEFEF} 
{ +\baby}                                       & {\textbf{0.6501}} & {\textbf{+0.74\%}} & { \textbf{0.8242}} & { \textbf{0.8256}} & { \textbf{0.8730}} & { \textbf{0.8475}}  & { \textbf{+0.40\%}}                                        & { \textbf{0.7969}} & { \textbf{0.8448}} & { \textbf{0.9093}}  & \multicolumn{1}{c|}{\cellcolor[HTML]{EFEFEF}{ \textbf{+0.26\%}}}                                                    & { \textbf{0.8473}} & { \textbf{0.8509}} & { \textbf{0.8458}}  & { \textbf{+0.55\%}}                                                    \\ \hline
\multicolumn{1}{l|}{{ CAN}}                                              & {0.6498} & {+0.54\%} & { 0.8227} & { 0.8238} & { 0.8721} & { 0.8464}  & { +0.08\%}                                        & { 0.7954} & { 0.8423} & { 0.9084}  & \multicolumn{1}{c|}{{ +0.04\%}}                                                    & { 0.8459} & { 0.8496} & { 0.8443}  & { +0.11\%}                                                    \\
 
{ +FRNet}                                        & {0.6500} & {+0.67\%} & { 0.8234} & { 0.8248} & { 0.8725} & { 0.8469}  & { +0.23\%}                                        & { 0.7958} & { 0.8429} & { 0.9086}  & \multicolumn{1}{c|}{{ +0.09\%}}                                                    & { 0.8466} & { 0.8503} & { 0.8447}  & { +0.23\%}                                                    \\
 
{ +PEPNet}                                       & {0.6490} & {+0.00\%} & { 0.8231} & { 0.8242} & { 0.8723} & { 0.8467}  & { +0.17\%}                                        & { 0.7958} & { 0.8427} & { 0.9085}  & \multicolumn{1}{c|}{{ +0.07\%}}                                                    & { 0.8463} & { 0.8502} & { 0.8450}  & { +0.31\%}                                                    \\
\rowcolor[HTML]{EFEFEF} 
{ +\baby}                                       & {\textbf{0.6503}} & {\textbf{+0.87\%}} & { \textbf{0.8240}} & { \textbf{0.8256}} & { \textbf{0.8731}} & { \textbf{0.8474}}  & { \textbf{+0.37\%}}                                        & { \textbf{0.7968}} & { \textbf{0.8446}} & { \textbf{0.9092}}  & \multicolumn{1}{c|}{\cellcolor[HTML]{EFEFEF}{ \textbf{+0.24\%}}}                                                    & { \textbf{0.8470}} & { \textbf{0.8509}} & { \textbf{0.8456}}  & { \textbf{+0.49\%}}                                                    \\ \hline
\end{tabular}
\end{adjustbox}
\label{tab:res}
\end{table*}

\subsection{Experimental Settings}

\paratitle{Datasets and Experimental Setup.} To sufficiently evaluate the proposed \baby, we conducts experiments on both a public dataset, i.e., Taobao Dataset \footnote{https://tianchi.aliyun.com/dataset/dataDetail?dataId=649} (\textit{Taobao}), and two \textbf{billion-scale} industrial dataset from Douyin, i.e., Douyin advertising platform (\textit{Douyin Ads}) and Douyin E-commerce platform (\textit{Douyin Ecom}). The statistics of the three datasets are reported in Table~\ref{tab:stats}.

\begin{itemize}[leftmargin=*]
    \item \textbf{Taobao}. The Taobao dataset released in \cite{zhu2018learning} provides user behavior data in Taobao and is currently widely used in sequential modeling methods \cite{cao2022sampling}. The former 7 days are used for training and the rest are for testing. Users with at least 200 interactions and 10 positive actions are filtered, and items with at least 10 interactions are filtered. There are 9,439 categories of items in the dataset, and we regard each category as a domain. In this dataset, page view is considered as a negative interaction and the other actions are regarded as positive label (order).
    
    \item \textbf{Douyin Ads}. We select the Conversion Rate (CVR) prediction task in Douyin Ads, and collect a subset of online traffic logs from Dec. 14th, 2022 to Mar. 10th, 2023, 87 days and 1.73 billion samples in total. The former 77 days are used for model training and the rest 10 days are used for evaluation. In Douyin Ads platform, according to different user external actions, the dataset can be divided into three major domains including pay in live, order in live, and shopping in short-videos, denoted by domain 1, 2, and 3, respectively.
    
    \item \textbf{Douyin Ecom}. Two kinds of user shopping behaviors including click and order are selected as the prediction targets in the Ecom service in Douyin Live, the most influential scenario in E-commerce services at ByteDance. A subset of online traffic logs from Jan. 1st to Mar. 1st, 2024 is collected, including 61 days and 2.52 billion samples. The first 54 days are selected for training and the last week is for validation. The two typical senarios in Douyin-Live, i.e., Live-Preview and Live-Slide, are involved in the dataset, denoted by domain 1 and 2, respectively.
\end{itemize}

\paratitle{Comparing methods and evaluation metrics.} To comprehensively compare the proposed \baby with existing methods, we select several representative SOTA models, which can be grouped into two categories: (1) the DNN-based methods including \textbf{DNN}, \textbf{DeepFM} \cite{guo2017deepfm}, \textbf{DCNv2} \cite{wang2021dcn}, \textbf{APG} \cite{yan2022apg}, \textbf{AdaSparse} \cite{yang2022adasparse}, \textbf{DFFM} \cite{guo2023dffm}, \textbf{MaskNet} \cite{wang2021masknet}; (2) the target attention-based backbone methods including \textbf{DIN} \cite{zhou2018deep}, \textbf{MHA} \cite{vaswani2017attention}, and \textbf{CAN} \cite{bian2022can}, based on which recent multi-domain embedding learning methods are added for comparison, including \textbf{FRNet} \cite{wang2022frnet} and \textbf{PEPNet} \cite{chang2023pepnet}. All the methods are implemented based on Tensorflow\cite{abadi2016tensorflow}, optimized by the cross-entropy loss. Adam \cite{kingma2014adam} optimizer is adopted with an initial learning rate of 0.00002. AUC metric is used to evaluate the ranking model performance. Further, we introduce relative improvement (Imp.) \cite{yan2014coupled} to measure the relative AUC gain, which is calculated as following as a random strategy yields AUC value at 0.5:

\begin{align}
AUC\ Imp. = \left(  \frac{AUC(Measured Model) - 0.5}{AUC(Baseline Model) - 0.5} - 1 \right) \times 100\%
\label{eqn:relaimp}
\end{align}

\subsection{Experimental Results}
\textbf{Overall Performance.} The comparison results of different methods on the three datasets are presented in Table~\ref{tab:res}. For Taobao dataset, only the overall results are provided due to the domain amount. For clarity, the results are reported with the comparing methods grouped into four groups, in which first group lists the DNN-based methods, and the rest-three groups list the multi-domain target attention methods based on DIN, MHA, and CAN, respectively. There are several observations from the results.

\textbf{First, it is observed that compared with the DNN-based approches, target-attention based sequential-modeling plays a vital role in ranking models.} From the table, it is demonstrated that DIN, MHA, and CAN achieves 0.54\% improvements in Taobao, 0.23\%, 0.17\%, and 0.08\% improvements in Douyin Ads, 0.22\%, 0.19\%, and 0.04\% improvements in click prediction, and 0.49\%, 0.31\%, and 0.11\% improvements in order prediction tasks in Douyin Ecom, respectively, demonstrating significant improvements.

\textbf{Second, the existing multi-domain methods contribute a positive effect for ranking models in general.} Specifically, in DNN-based methods, it is observed that AdaSparse outperforms the baseline in Taobao and Douyin Ads. FRNet and PEPNet also show improved performance in different groups.

\textbf{Finally, the proposed \baby consistently achieves the best performance in different groups with DIN, MHA, and CAN as backbone, showing its high effectiveness and compatibility.} Specifically, in Taobao, the proposed \baby outperforms other SOTAs with 0.47\%, 0.20\%, and 0.20\% with the second-best methods in the DIN-, MHA-, and CAN-based groups. In Douyin Ads, \baby beats other methods and achieves 0.20\%, 0.12\%, and 0.14\% improvements compared with the second-best approach. In Douyin Ecom, \baby outperforms the second-best approach in the three groups by 0.05\%, 0.07\%, and 0.15\% in click prediction task, and by 0.11\%, 0.15\%, and 0.18\% in order prediction task, respectively. Besides, in terms of each domain of the two industrial datasets, \baby solidly outperforms the compared methods. Thus, the promising results show the superiority of the personalized target attention mechanism.

\begin{table}[]
\caption{Ablation study of the proposed \baby.}
\resizebox{0.48\textwidth}{!}{%
\begin{tabular}{c|c|ccccc}
\hline
Groups                                                                & Methods                                                   & D1     & D2     & D3     & Overall & \begin{tabular}[c]{@{}c@{}}Overall\\ Imp.\end{tabular} \\ \hline
\multirow{3}{*}{\begin{tabular}[c]{@{}c@{}} \\ DIN-\\ based\end{tabular}} & ADS                                                    & 0.8245 & 0.8262 & 0.8733 & 0.8477  & -                                                      \\
                                                                      & w/o PCRG                                                  & 0.8242 & 0.8257 & 0.8730 & 0.8475  & \textbf{-0.06\%}                                                \\
                                                                      & \begin{tabular}[c]{@{}c@{}}w/o PCRG\\ \&PSRG\end{tabular} & 0.8235 & 0.8251 & 0.8724 & 0.8469  & \textbf{-0.23\%}                                                \\ \hline
\multirow{3}{*}{\begin{tabular}[c]{@{}c@{}} \\MHA-\\ based\end{tabular}} & ADS                                                    & 0.8242 & 0.8256 & 0.8730 & 0.8475  & -                                                      \\
                                                                      & w/o PCRG                                                  & 0.8241 & 0.8252 & 0.8730 & 0.8474  & \textbf{-0.03\%}                                                \\
                                                                      & \begin{tabular}[c]{@{}c@{}}w/o PCRG\\ \&PSRG\end{tabular} & 0.8232 & 0.8238 & 0.8721 & 0.8467  & \textbf{-0.23\%}                                                \\ \hline
\multirow{3}{*}{\begin{tabular}[c]{@{}c@{}} \\CAN-\\ based\end{tabular}} & ADS                                                    & 0.8240 & 0.8256 & 0.8731 & 0.8474  & -                                                      \\
                                                                      & w/o PCRG                                                  & 0.8230 & 0.8240 & 0.8723 & 0.8466  & \textbf{-0.23\%}                                                \\
                                                                      & \begin{tabular}[c]{@{}c@{}}w/o PCRG\\ \&PSRG\end{tabular} & 0.8227 & 0.8238 & 0.8721 & 0.8464  & \textbf{-0.29\%}                                                \\ \hline
\end{tabular}
}
\label{tab:ablation}
\end{table}

\subsection{Ablation Study \& Sensitivity Analysis}
\textbf{Ablation study}. To further evaluate the performance of the two modules in \baby, i.e., PCRG and PSRG, here we perform ablation study in the challenging Douyin Ads dataset. As shown in Table~\ref{tab:ablation}, after ablating the PCRG module, the overall performance is dropped by 0.06\%, 0.03\%, and 0.23\% in DIN-, MHA-, and CAN-based methods. Besides, after ablating both the PCRG and PSRG modules, the overall performance is decreased by 0.23\%, 0.23\%, and 0.29\%, respectively. It can thus be concluded that both the personalized target item and the personalized sequences contributes positive impact to the proposed \baby, confirming the validity of these modules.


\begin{figure}[h]
\center
\includegraphics[width=\linewidth]{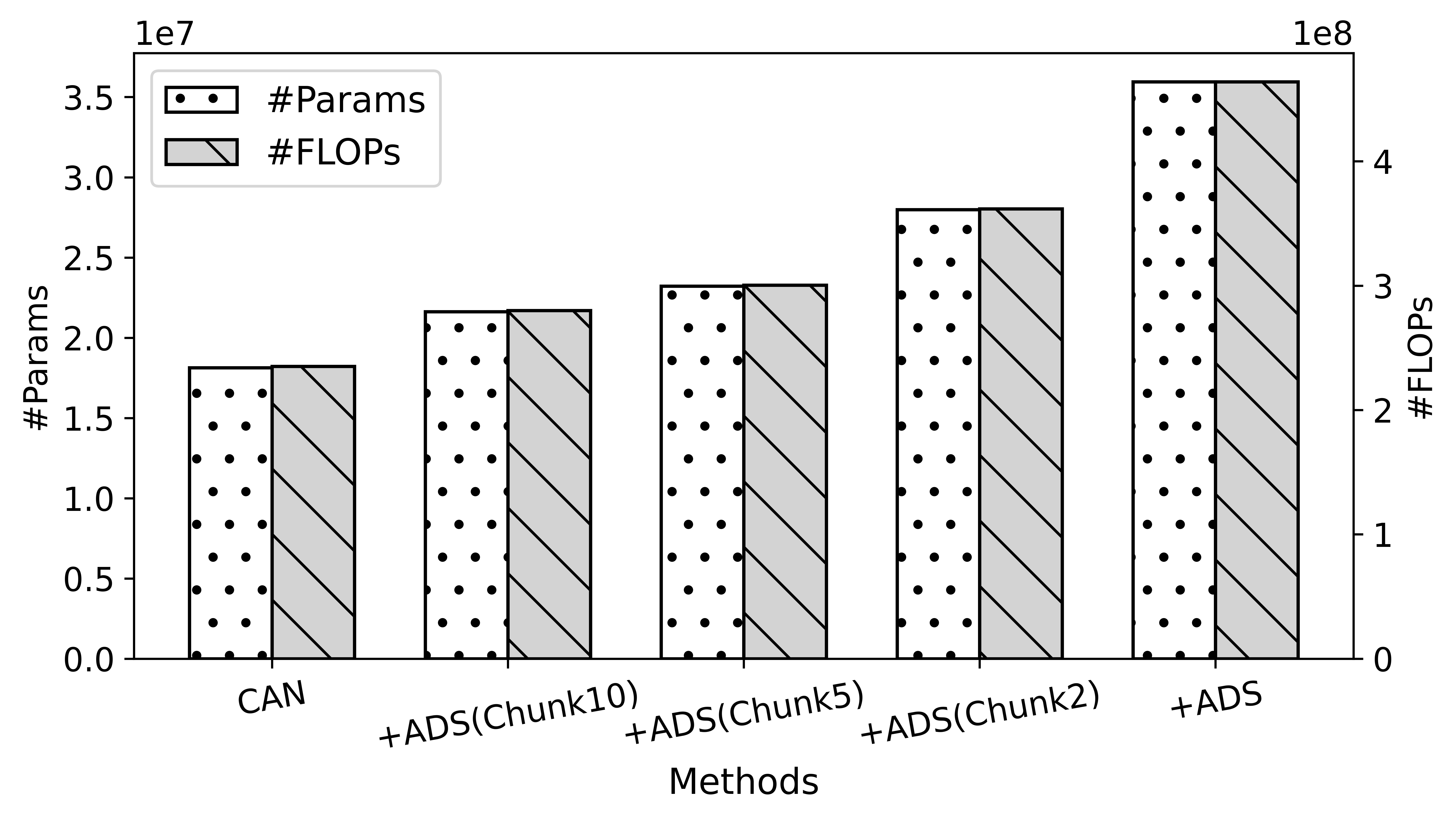}
\caption{Model parameter and training FLOPs patterns by varing the number of chunks in \baby, where "Chunk \textit{K}" means there are \textit{K} items grouped into a chunk in PCRG.}
\label{fig:sensitivity_efficiency}
\end{figure}

\begin{figure*}[t]
\captionsetup[subfigure]{labelformat=empty}
\centering
\begin{minipage}[b]{\linewidth}
\subfigure{
 \includegraphics[width=0.16\linewidth,trim=8.2 10 9 7.2,clip]{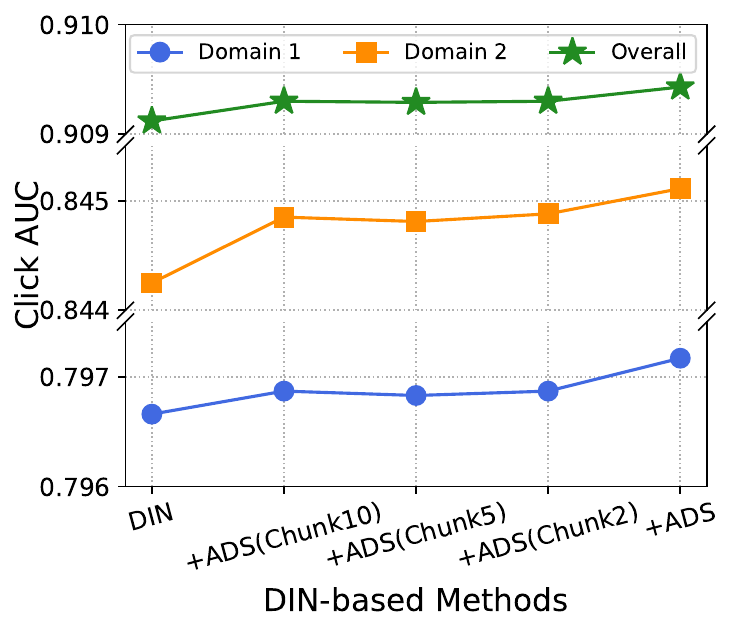}
}
\hspace{-0.22cm}
\subfigure{
\includegraphics[width=0.16\linewidth,trim=8.2 10 9 7.2,clip]{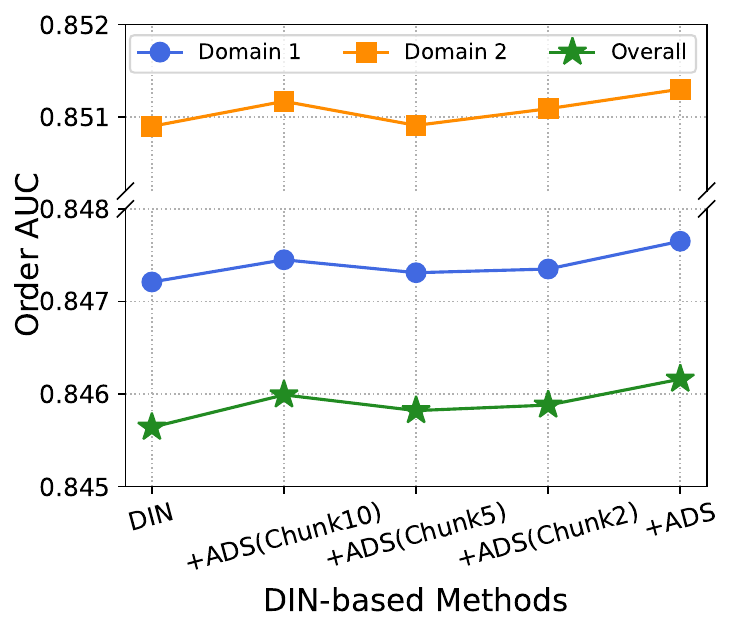}
}
\hspace{-0.22cm}
\subfigure{
\includegraphics[width=0.16\linewidth,trim=8.2 10 9 7.2,clip]{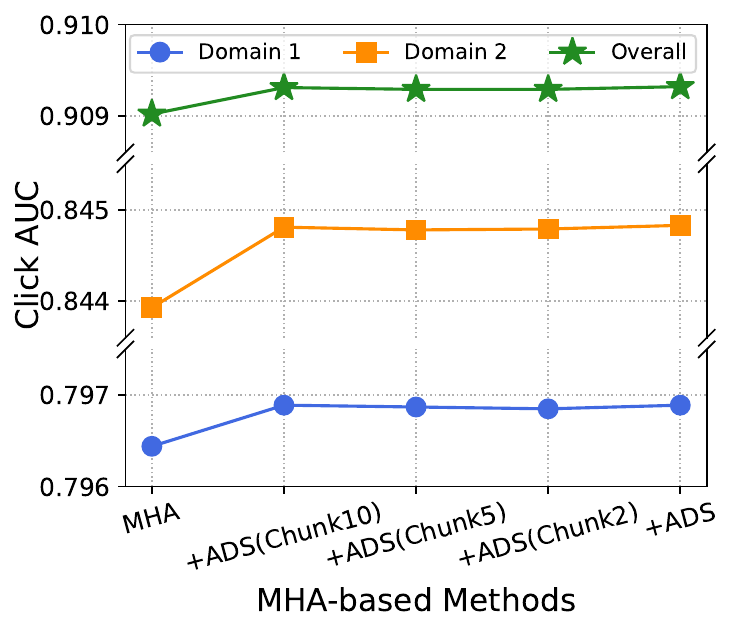}
}
\hspace{-0.22cm}
\subfigure{
\includegraphics[width=0.16\linewidth,trim=8.2 10 9 7.2,clip]{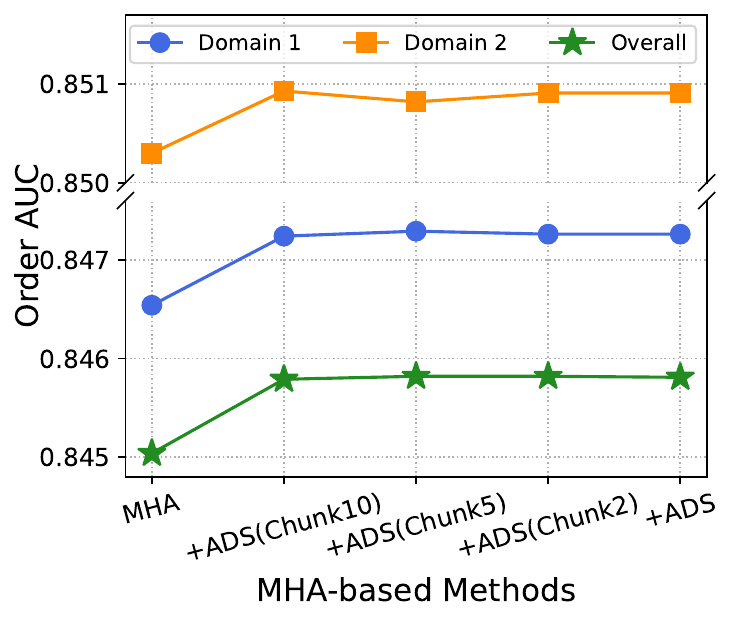}
}
\hspace{-0.22cm}
\subfigure{
\includegraphics[width=0.16\linewidth,trim=8.2 10 9 7.2,clip]{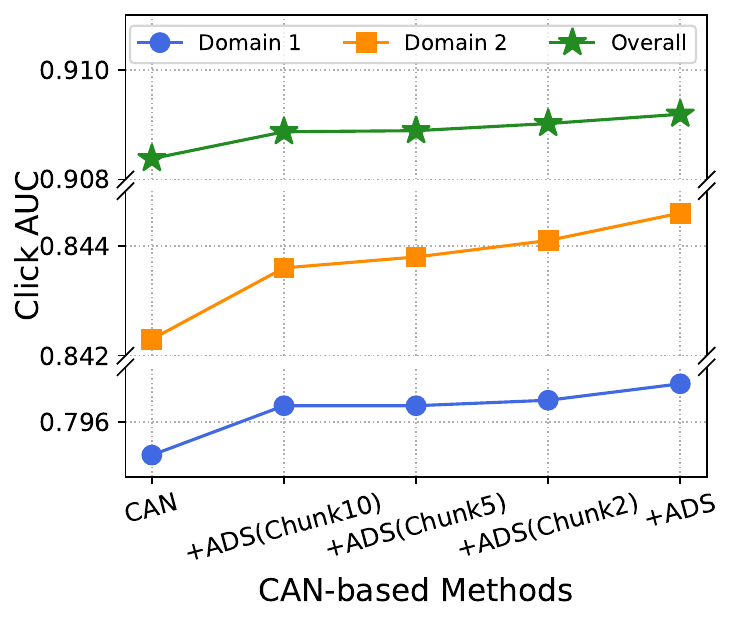}
}
\hspace{-0.22cm}
\subfigure{
\includegraphics[width=0.16\linewidth,trim=8.2 10 9 7.2,clip]{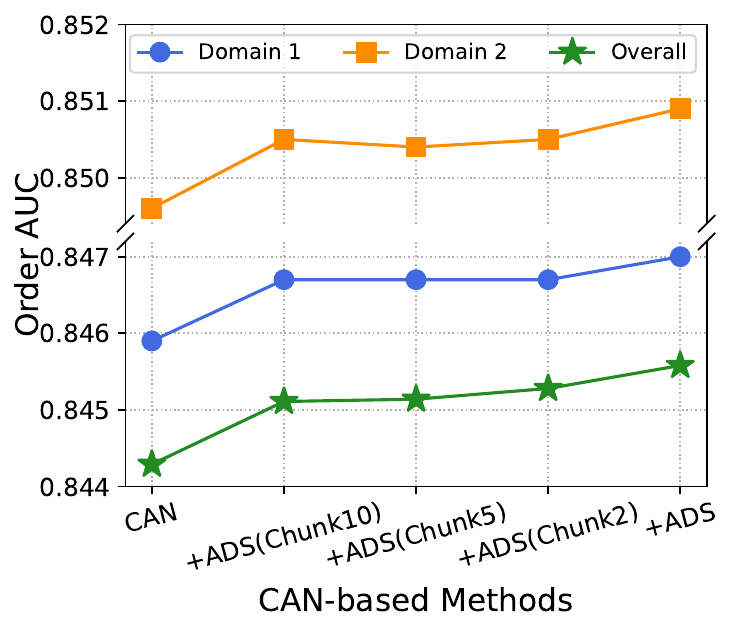}
}
\end{minipage}
\caption{Performance patterns by varying the number of chunks in \baby.}
\label{fig:sensitivity_performance}
\end{figure*}

\textbf{Sensitivity Analysis to Number of Chunks in \baby.} 
Furthermore, to investigate the impact of number of chunks, sensitivity analysis for the \baby are conducted. Specifically, we investigate the performance patterns by varying the number of items in each chunk from [1, 2, 5, 10] in Douyin E-commerce, and the sensitivity analysis is performed from two aspects, i.e., training efficiency and model performance.
\begin{itemize}[leftmargin=*]

    \item \textbf{Training efficiency patterns.} We evaluate the model training efficiency by summarizing and comparing the model parameters and training floating point operations (FLOPs) under different chunks in \baby, and the results are illustrated in Fig~\ref{fig:sensitivity_efficiency}. From the figure, it can be clearly observed that with the modeling becomes more personalized, the model parameters and training FLOPs consistently increase.
    
    \item \textbf{Performance patterns.} The model performance pattern by varying the number of items in each chunk is illustrated in Fig~\ref{fig:sensitivity_performance}. Specifically, first, in comparison with the vanilla DIN, MHA, and CAN, \baby and \baby with different chunks show obvious performance improvement in terms of both click and order prediction tasks. Besides, it can be observed that with number of items covered in the chunk decreases, the model performance continues to increase, and the most personalized model, i.e., \baby without chunk, achieves the best performance, showing that it is of significance to consider the personality characterization of the candidate item.

\end{itemize}
Overall, both the performance and the training cost increase with the personalization ability in \baby increased, while we observe that even with a small increase in training cost (\baby with Chunk 10 compared with vanilla method), the model performance still achieves promising gains, and thus practitioners can select the parameters with more flexibility according to the balance of effectiveness and efficiency.

\begin{table}[]
\caption{Online A/B results in Douyin Ads. The boldface highlight denotes that the improvement is significant with $p < $ \textbf{0.01}.}

\begin{tabular}{c|ccc}
\hline
                                                                                    & Methods  & CPM              & ADVV             \\ \hline
\multirow{2}{*}{\begin{tabular}[c]{@{}c@{}}Domain 1\\ (Short-Video)\end{tabular}}   & baseline & +0.00\%          & +0.00\%          \\
                                                                                    & \baby    & \textbf{+1.01\%} & \textbf{+1.66\%} \\ \hline
\multirow{2}{*}{\begin{tabular}[c]{@{}c@{}}Domain 2\\ (Live)\end{tabular}} & baseline & +0.00\%          & +0.00\%          \\
                                                                                    & \baby    & \textbf{+0.39\%} & \textbf{+0.84\%} \\ \hline
\multirow{2}{*}{Overall}                                                            & baseline & +0.00\%          & +0.00\%          \\
                                                                                    & \baby    & \textbf{+0.52\%} & \textbf{+1.00\%} \\ \hline
\end{tabular}
\label{tab:ab_ads}
\end{table}


\begin{table}[]
\caption{Online A/B results in Douyin Ecom. The boldface highlight denotes that the improvement is significant with $p < $ \textbf{0.01}.}
\begin{tabular}{c|cccc}
\hline
                                                                                   & Methods  & GMV/U            & Order/U          & GPM              \\ \hline
\multirow{2}{*}{\begin{tabular}[c]{@{}c@{}}Domain 1\\ (Live-Preview)\end{tabular}} & baseline & +0.00\%          & +0.00\%          & +0.00\%          \\
                                                                                   & \baby    & \textbf{+0.69\%} & \textbf{+0.32\%} & \textbf{+0.78\%} \\ \hline
\multirow{2}{*}{\begin{tabular}[c]{@{}c@{}}Domain 2\\ (Live-Slide)\end{tabular}}   & baseline & +0.00\%          & +0.00\%          & +0.00\%          \\
                                                                                   & \baby    & \textbf{+0.93\%} & \textbf{+0.36\%} & \textbf{+0.97\%} \\ \hline
\multirow{2}{*}{Overall}                                                           & baseline & +0.00\%          & +0.00\%          & +0.00\%          \\
                                                                                   & \baby    & \textbf{+0.79\%} & \textbf{+0.36\%} & \textbf{+0.89\%} \\ \hline
\end{tabular}
\label{tab:ab_ecom}
\end{table}

\subsection{Online Deployments}

The \baby model is online deployed with distribution across multi GPUs with a sharding and data parallelism strategy. The low-frequency embeddings are eliminated to reduce storage. To further increase GPU throughputs, we introduce Dense Computation Asynchrony strategy. It splits the computation graph into SparseForward and DenseCompute parts, which enables a pipeline effect and greatly improves training and inference efficiency. Benefiting from these, the offline training resource remains the same as baseline, i.e., 64 Nvidia A100s. Take the Douyin Ads as an example, the training time cost slightly increases from 41.3 to 42.8 hours (+3.6\%). The online latency remains at 30ms with no significant changes.

\subsection{Online A/B Experiments}

To investigate the performance of the proposed \baby in real industrial scenarios, we perform careful online A/B testing in both the advertising system and the e-commerce system in Douyin, respectively. 

\begin{itemize}[leftmargin=*]


\item \textbf{Douyin Ads}. The online experiment on Douyin Ads is conducted from Nov. 2nd to Nov. 8th, 2023, hitting 74,079,729 users in Douyin APP. Two metrics including Cost Per Mile (CPM) and Advertiser Value (ADVV) are selected for comparison. Note that the deployed scenario serves as the major traffic for advertisement in ByteDance with a strong baseline, where a 0.5\% improvement in ADVV or CPM is considered to be significant. The comparison results are shown in Table~\ref{tab:ab_ads}. The domains including pay in live and order in live are summarized as Domain 2 (Live). From the table, it is observed that after deploying the \baby, the overall CPM is improved by 0.52\%, and the ADVV is improved by 1.00\%, showing the merits of adaptive sequence modeling of the proposed \baby. Besides, in the two major domains in Douyin Ads, including both live and short-video, the proposed \baby beats the baseline and achieves consistent improvements, showing the effectiveness in domain-aware sequential modeling for large-scaled industrial recommenders.


\item \textbf{Douyin E-commerce}. The online experiment is performed from Jan. 23rd to Jan. 29th, 2024, on Douyin Ecom, hitting 508,926,918 users in Douyin APP. The experimental results are presented in Table~\ref{tab:ab_ecom}. Three metrics are selected for comparison, i.e., the gross merchandise volume per user (GMV/U), the number of orders per user (Order/U), and GMV per Mille (GPM), which are all important commercial metrics in Douyin E-commerce. Similar to the experiment conducted on Douyin Ads, it is noted that this deployed scenario contributes the highest GMV in ByteDance with a very strong baseline, and generally, an improvement at the ratio of 0.5\% in GMV is considered to be significant. As shown in Table~\ref{tab:ab_ecom}, the overall GMV/U, Order/U, and GPM are uplifted by 0.79\%, 0.36\%, and 0.89\%, respectively. Besides, consistent improvements are observed in both domains including live-preview and live-slide, and all the improvements are tested to be substantially statistically significant with a $p-\rm{value} < $ 0.01, verifying its efficacy.

\end{itemize}

\section{Related Works}
\subsection{Ranking models in recommendation}
The ranking model generally serves as the last stage of the industrial recommendation system funnel and plays a pivotal role in personalizing user recommendations in content and e-commerce platforms \cite{covington2016deep, chai2022umi}. Typically, they can be divided into two categories: 1) Feature interaction models which focus on modeling  discrete identifiers (IDs) feature interactions and capturing the co-occurrence patterns \cite{rendle2010factorization, guo2017deepfm, cheng2016wide, wang2017deep, wang2021dcn, zhang2022dhen}, and 2) Target-aware sequential models which capture user intention by evaluating the attention weight between the candidate item and the behavioral items \cite{zhou2018deep, zhou2019deep, shen2022deep, pi2020search, bian2022can, vaswani2017attention}. Despite the broad developments, the multi-domain problem that widely exists in large-scale recommender systems has been rarely considered in these methods.

\subsection{Multi-Domain Recommendation}
Multi-domain recommendation aims to provide accurate personalized recommendation results for users and items from multiple domains. Both coarse-grained \cite{sheng2021one, tang2020progressive} and fine-grained \cite{yan2022apg, yang2022adasparse} multi-domain methods have been developed for sophisticated network structures, while the multi-domain learning for sequential modeling is seldom considered. Some other methods also contribute to learning personalized embeddings for multi-domain recommendation. The early idea is to use the self-gate mechanism to generate bitwise weights to reweigh the embeddings \cite{huang2020gatenet}. Further, the feature refinement network (FRNet) \cite{wang2022frnet} takes user contexts as gate input, enabling user-level personalization on specific features. To further consider the impact of multidomains, parameter and embedding personalized network (PEPNet) \cite{chang2023pepnet} and domain-facilitated feature modeling (DFFM) \cite{guo2023dffm} are developed, while bitwise scaling in PEPNet and linear transformation in DFFM are not flexible and sufficient enough to capture multidomains. Differently, \baby provides an approach by transforming the sequence embedding through a generated meta network, which provides a more flexible share-and-private modeling structure and is validated to be superior to existing methods. Besides, the \baby highlights that the identical target item has different influences on different sequence items. Traditional works can only obtain a novel target embedding shared for all sequence items, rather than multiple queries for different items, and this effective approach has not been explored in recommendation, to our knowledge. Besides, the extensive experiments on both public dataset and two giant scenarios in ByteDance (billions of daily active users) have demonstrated the effectiveness compared with existing approaches.

\section{Conclusion}
In this paper, we propose a novel multi-domain ranking model named adaptive domain scaling (ADS) model, which builds personalized sequence representation generation (PSRG) and personalized candidate representation generation (PCRG) to separately generate personalized sequence item and candidate target item representations. Specifically, for PSRG, both a sequence-weight gen-net and sequence-bias gen-net are developed to formulate the instance-wise MLP and modify the sequence items, such that the identical item occured in different user sequences has diverse representation. Besides, in PCRG, a multi-query gen-net is devised to generate multiple queries for sequence items, such that the identical candidate item extracts diverse intentions of users from the sequences from a multi-domain perspective. Offline experiments on both a public and two industrial datasets validate its significant and consistent improvements over existing SOTA methods, and extensive online experiments on two influential scenarios in ByteDance demonstrate its effectiveness. 

\baby is currently fully deployed and achieves substantial enhancements in dozens of recommendation services at ByteDance, and now serves billions of users each day. We believe that it has offered a solid solution and propelled the advancements in multi-domain ranking from a sequential-modeling perspective, and more efficient and effective adaptive modeling structures will be explored in the future.


\balance
\bibliographystyle{ACM-Reference-Format}
\bibliography{ref}


\begin{thebibliography}{34}


\ifx \showCODEN    \undefined \def \showCODEN     #1{\unskip}     \fi
\ifx \showDOI      \undefined \def \showDOI       #1{#1}\fi
\ifx \showISBNx    \undefined \def \showISBNx     #1{\unskip}     \fi
\ifx \showISBNxiii \undefined \def \showISBNxiii  #1{\unskip}     \fi
\ifx \showISSN     \undefined \def \showISSN      #1{\unskip}     \fi
\ifx \showLCCN     \undefined \def \showLCCN      #1{\unskip}     \fi
\ifx \shownote     \undefined \def \shownote      #1{#1}          \fi
\ifx \showarticletitle \undefined \def \showarticletitle #1{#1}   \fi
\ifx \showURL      \undefined \def \showURL       {\relax}        \fi
\providecommand\bibfield[2]{#2}
\providecommand\bibinfo[2]{#2}
\providecommand\natexlab[1]{#1}
\providecommand\showeprint[2][]{arXiv:#2}

\bibitem[Abadi et~al\mbox{.}(2016)]%
        {abadi2016tensorflow}
\bibfield{author}{\bibinfo{person}{Mart{\'\i}n Abadi}, \bibinfo{person}{Paul Barham}, \bibinfo{person}{Jianmin Chen}, \bibinfo{person}{Zhifeng Chen}, \bibinfo{person}{Andy Davis}, \bibinfo{person}{Jeffrey Dean}, \bibinfo{person}{Matthieu Devin}, \bibinfo{person}{Sanjay Ghemawat}, \bibinfo{person}{Geoffrey Irving}, \bibinfo{person}{Michael Isard}, {et~al\mbox{.}}} \bibinfo{year}{2016}\natexlab{}.
\newblock \showarticletitle{TensorFlow: a system for large-scale machine learning}. In \bibinfo{booktitle}{\emph{12th USENIX symposium on operating systems design and implementation (OSDI 16)}}. \bibinfo{pages}{265--283}.
\newblock


\bibitem[Bian et~al\mbox{.}(2022)]%
        {bian2022can}
\bibfield{author}{\bibinfo{person}{Weijie Bian}, \bibinfo{person}{Kailun Wu}, \bibinfo{person}{Lejian Ren}, \bibinfo{person}{Qi Pi}, \bibinfo{person}{Yujing Zhang}, \bibinfo{person}{Can Xiao}, \bibinfo{person}{Xiang-Rong Sheng}, \bibinfo{person}{Yong-Nan Zhu}, \bibinfo{person}{Zhangming Chan}, \bibinfo{person}{Na Mou}, {et~al\mbox{.}}} \bibinfo{year}{2022}\natexlab{}.
\newblock \showarticletitle{CAN: feature co-action network for click-through rate prediction}. In \bibinfo{booktitle}{\emph{Proceedings of the fifteenth ACM international conference on web search and data mining}}. \bibinfo{pages}{57--65}.
\newblock


\bibitem[Cao et~al\mbox{.}(2022)]%
        {cao2022sampling}
\bibfield{author}{\bibinfo{person}{Yue Cao}, \bibinfo{person}{Xiaojiang Zhou}, \bibinfo{person}{Jiaqi Feng}, \bibinfo{person}{Peihao Huang}, \bibinfo{person}{Yao Xiao}, \bibinfo{person}{Dayao Chen}, {and} \bibinfo{person}{Sheng Chen}.} \bibinfo{year}{2022}\natexlab{}.
\newblock \showarticletitle{Sampling is all you need on modeling long-term user behaviors for CTR prediction}. In \bibinfo{booktitle}{\emph{Proceedings of the 31st ACM International Conference on Information \& Knowledge Management}}. \bibinfo{pages}{2974--2983}.
\newblock


\bibitem[Caruana(1993)]%
        {caruana1993multitask}
\bibfield{author}{\bibinfo{person}{R Caruana}.} \bibinfo{year}{1993}\natexlab{}.
\newblock \showarticletitle{Multitask learning: A knowledge-based source of inductive bias1}. In \bibinfo{booktitle}{\emph{Proceedings of the Tenth International Conference on Machine Learning}}. Citeseer, \bibinfo{pages}{41--48}.
\newblock


\bibitem[Chai et~al\mbox{.}(2022)]%
        {chai2022umi}
\bibfield{author}{\bibinfo{person}{Zheng Chai}, \bibinfo{person}{Zhihong Chen}, \bibinfo{person}{Chenliang Li}, \bibinfo{person}{Rong Xiao}, \bibinfo{person}{Houyi Li}, \bibinfo{person}{Jiawei Wu}, \bibinfo{person}{Jingxu Chen}, {and} \bibinfo{person}{Haihong Tang}.} \bibinfo{year}{2022}\natexlab{}.
\newblock \showarticletitle{User-aware multi-interest learning for candidate matching in recommenders}. In \bibinfo{booktitle}{\emph{Proceedings of the 45th International ACM SIGIR Conference on Research and Development in Information Retrieval}}. \bibinfo{pages}{1326--1335}.
\newblock


\bibitem[Chang et~al\mbox{.}(2023)]%
        {chang2023pepnet}
\bibfield{author}{\bibinfo{person}{Jianxin Chang}, \bibinfo{person}{Chenbin Zhang}, \bibinfo{person}{Yiqun Hui}, \bibinfo{person}{Dewei Leng}, \bibinfo{person}{Yanan Niu}, \bibinfo{person}{Yang Song}, {and} \bibinfo{person}{Kun Gai}.} \bibinfo{year}{2023}\natexlab{}.
\newblock \showarticletitle{Pepnet: Parameter and embedding personalized network for infusing with personalized prior information}. In \bibinfo{booktitle}{\emph{Proceedings of the 29th ACM SIGKDD Conference on Knowledge Discovery and Data Mining}}. \bibinfo{pages}{3795--3804}.
\newblock


\bibitem[Cheng et~al\mbox{.}(2016)]%
        {cheng2016wide}
\bibfield{author}{\bibinfo{person}{Heng-Tze Cheng}, \bibinfo{person}{Levent Koc}, \bibinfo{person}{Jeremiah Harmsen}, \bibinfo{person}{Tal Shaked}, \bibinfo{person}{Tushar Chandra}, \bibinfo{person}{Hrishi Aradhye}, \bibinfo{person}{Glen Anderson}, \bibinfo{person}{Greg Corrado}, \bibinfo{person}{Wei Chai}, \bibinfo{person}{Mustafa Ispir}, {et~al\mbox{.}}} \bibinfo{year}{2016}\natexlab{}.
\newblock \showarticletitle{Wide \& deep learning for recommender systems}. In \bibinfo{booktitle}{\emph{Proceedings of the 1st workshop on deep learning for recommender systems}}. \bibinfo{pages}{7--10}.
\newblock


\bibitem[Covington et~al\mbox{.}(2016)]%
        {covington2016deep}
\bibfield{author}{\bibinfo{person}{Paul Covington}, \bibinfo{person}{Jay Adams}, {and} \bibinfo{person}{Emre Sargin}.} \bibinfo{year}{2016}\natexlab{}.
\newblock \showarticletitle{Deep neural networks for youtube recommendations}. In \bibinfo{booktitle}{\emph{Proceedings of the 10th ACM conference on recommender systems}}. \bibinfo{pages}{191--198}.
\newblock


\bibitem[Guo et~al\mbox{.}(2017)]%
        {guo2017deepfm}
\bibfield{author}{\bibinfo{person}{Huifeng Guo}, \bibinfo{person}{Ruiming Tang}, \bibinfo{person}{Yunming Ye}, \bibinfo{person}{Zhenguo Li}, {and} \bibinfo{person}{Xiuqiang He}.} \bibinfo{year}{2017}\natexlab{}.
\newblock \showarticletitle{DeepFM: a factorization-machine based neural network for CTR prediction}. In \bibinfo{booktitle}{\emph{Proceedings of the 26th International Joint Conference on Artificial Intelligence}}. \bibinfo{pages}{1725--1731}.
\newblock


\bibitem[Guo et~al\mbox{.}(2023)]%
        {guo2023dffm}
\bibfield{author}{\bibinfo{person}{Wei Guo}, \bibinfo{person}{Chenxu Zhu}, \bibinfo{person}{Fan Yan}, \bibinfo{person}{Bo Chen}, \bibinfo{person}{Weiwen Liu}, \bibinfo{person}{Huifeng Guo}, \bibinfo{person}{Hongkun Zheng}, \bibinfo{person}{Yong Liu}, {and} \bibinfo{person}{Ruiming Tang}.} \bibinfo{year}{2023}\natexlab{}.
\newblock \showarticletitle{DFFM: Domain Facilitated Feature Modeling for CTR Prediction}. In \bibinfo{booktitle}{\emph{Proceedings of the 32nd ACM International Conference on Information and Knowledge Management}}. \bibinfo{pages}{4602--4608}.
\newblock


\bibitem[Huang et~al\mbox{.}(2020)]%
        {huang2020gatenet}
\bibfield{author}{\bibinfo{person}{Tongwen Huang}, \bibinfo{person}{Qingyun She}, \bibinfo{person}{Zhiqiang Wang}, {and} \bibinfo{person}{Junlin Zhang}.} \bibinfo{year}{2020}\natexlab{}.
\newblock \showarticletitle{GateNet: gating-enhanced deep network for click-through rate prediction}.
\newblock \bibinfo{journal}{\emph{arXiv preprint arXiv:2007.03519}} (\bibinfo{year}{2020}).
\newblock


\bibitem[Jiang et~al\mbox{.}(2022)]%
        {jiang2022adaptive}
\bibfield{author}{\bibinfo{person}{Yuchen Jiang}, \bibinfo{person}{Qi Li}, \bibinfo{person}{Han Zhu}, \bibinfo{person}{Jinbei Yu}, \bibinfo{person}{Jin Li}, \bibinfo{person}{Ziru Xu}, \bibinfo{person}{Huihui Dong}, {and} \bibinfo{person}{Bo Zheng}.} \bibinfo{year}{2022}\natexlab{}.
\newblock \showarticletitle{Adaptive domain interest network for multi-domain recommendation}. In \bibinfo{booktitle}{\emph{Proceedings of the 31st ACM International Conference on Information \& Knowledge Management}}. \bibinfo{pages}{3212--3221}.
\newblock


\bibitem[Kingma and Ba(2015)]%
        {kingma2014adam}
\bibfield{author}{\bibinfo{person}{Diederik~P Kingma} {and} \bibinfo{person}{Jimmy Ba}.} \bibinfo{year}{2015}\natexlab{}.
\newblock \showarticletitle{Adam: A method for stochastic optimization}. In \bibinfo{booktitle}{\emph{Proceedings of ICLR}}.
\newblock


\bibitem[Li et~al\mbox{.}(2023)]%
        {li2023one}
\bibfield{author}{\bibinfo{person}{Chenglin Li}, \bibinfo{person}{Yuanzhen Xie}, \bibinfo{person}{Chenyun Yu}, \bibinfo{person}{Bo Hu}, \bibinfo{person}{Zang Li}, \bibinfo{person}{Guoqiang Shu}, \bibinfo{person}{Xiaohu Qie}, {and} \bibinfo{person}{Di Niu}.} \bibinfo{year}{2023}\natexlab{}.
\newblock \showarticletitle{One for all, all for one: Learning and transferring user embeddings for cross-domain recommendation}. In \bibinfo{booktitle}{\emph{Proceedings of the Sixteenth ACM International Conference on Web Search and Data Mining}}. \bibinfo{pages}{366--374}.
\newblock


\bibitem[Lu et~al\mbox{.}(2025)]%
        {lu2025large}
\bibfield{author}{\bibinfo{person}{Hui Lu}, \bibinfo{person}{Zheng Chai}, \bibinfo{person}{Yuchao Zheng}, \bibinfo{person}{Zhe Chen}, \bibinfo{person}{Deping Xie}, \bibinfo{person}{Peng Xu}, \bibinfo{person}{Xun Zhou}, {and} \bibinfo{person}{Di Wu}.} \bibinfo{year}{2025}\natexlab{}.
\newblock \showarticletitle{Large Memory Network for Recommendation}. In \bibinfo{booktitle}{\emph{Proceedings of the ACM Web Conference}}.
\newblock
\urldef\tempurl%
\url{https://doi.org/10.1145/3701716.3715514}
\showDOI{\tempurl}


\bibitem[Pi et~al\mbox{.}(2020)]%
        {pi2020search}
\bibfield{author}{\bibinfo{person}{Qi Pi}, \bibinfo{person}{Guorui Zhou}, \bibinfo{person}{Yujing Zhang}, \bibinfo{person}{Zhe Wang}, \bibinfo{person}{Lejian Ren}, \bibinfo{person}{Ying Fan}, \bibinfo{person}{Xiaoqiang Zhu}, {and} \bibinfo{person}{Kun Gai}.} \bibinfo{year}{2020}\natexlab{}.
\newblock \showarticletitle{Search-based user interest modeling with lifelong sequential behavior data for click-through rate prediction}. In \bibinfo{booktitle}{\emph{Proceedings of the 29th ACM International Conference on Information \& Knowledge Management}}. \bibinfo{pages}{2685--2692}.
\newblock


\bibitem[Rendle(2010)]%
        {rendle2010factorization}
\bibfield{author}{\bibinfo{person}{Steffen Rendle}.} \bibinfo{year}{2010}\natexlab{}.
\newblock \showarticletitle{Factorization machines}. In \bibinfo{booktitle}{\emph{2010 IEEE International conference on data mining}}. IEEE, \bibinfo{pages}{995--1000}.
\newblock


\bibitem[Shen et~al\mbox{.}(2022)]%
        {shen2022deep}
\bibfield{author}{\bibinfo{person}{Qijie Shen}, \bibinfo{person}{Hong Wen}, \bibinfo{person}{Wanjie Tao}, \bibinfo{person}{Jing Zhang}, \bibinfo{person}{Fuyu Lv}, \bibinfo{person}{Zulong Chen}, {and} \bibinfo{person}{Zhao Li}.} \bibinfo{year}{2022}\natexlab{}.
\newblock \showarticletitle{Deep interest highlight network for click-through rate prediction in trigger-induced recommendation}. In \bibinfo{booktitle}{\emph{Proceedings of the ACM Web Conference 2022}}. \bibinfo{pages}{422--430}.
\newblock


\bibitem[Sheng et~al\mbox{.}(2021)]%
        {sheng2021one}
\bibfield{author}{\bibinfo{person}{Xiang-Rong Sheng}, \bibinfo{person}{Liqin Zhao}, \bibinfo{person}{Guorui Zhou}, \bibinfo{person}{Xinyao Ding}, \bibinfo{person}{Binding Dai}, \bibinfo{person}{Qiang Luo}, \bibinfo{person}{Siran Yang}, \bibinfo{person}{Jingshan Lv}, \bibinfo{person}{Chi Zhang}, \bibinfo{person}{Hongbo Deng}, {et~al\mbox{.}}} \bibinfo{year}{2021}\natexlab{}.
\newblock \showarticletitle{One model to serve all: Star topology adaptive recommender for multi-domain ctr prediction}. In \bibinfo{booktitle}{\emph{Proceedings of the 30th ACM International Conference on Information \& Knowledge Management}}. \bibinfo{pages}{4104--4113}.
\newblock


\bibitem[Tang et~al\mbox{.}(2020)]%
        {tang2020progressive}
\bibfield{author}{\bibinfo{person}{Hongyan Tang}, \bibinfo{person}{Junning Liu}, \bibinfo{person}{Ming Zhao}, {and} \bibinfo{person}{Xudong Gong}.} \bibinfo{year}{2020}\natexlab{}.
\newblock \showarticletitle{Progressive layered extraction (ple): A novel multi-task learning (mtl) model for personalized recommendations}. In \bibinfo{booktitle}{\emph{Proceedings of the 14th ACM Conference on Recommender Systems}}. \bibinfo{pages}{269--278}.
\newblock


\bibitem[Vaswani et~al\mbox{.}(2017)]%
        {vaswani2017attention}
\bibfield{author}{\bibinfo{person}{Ashish Vaswani}, \bibinfo{person}{Noam Shazeer}, \bibinfo{person}{Niki Parmar}, \bibinfo{person}{Jakob Uszkoreit}, \bibinfo{person}{Llion Jones}, \bibinfo{person}{Aidan~N Gomez}, \bibinfo{person}{{\L}ukasz Kaiser}, {and} \bibinfo{person}{Illia Polosukhin}.} \bibinfo{year}{2017}\natexlab{}.
\newblock \showarticletitle{Attention is all you need}.
\newblock \bibinfo{journal}{\emph{Advances in neural information processing systems}}  \bibinfo{volume}{30} (\bibinfo{year}{2017}).
\newblock


\bibitem[Wang et~al\mbox{.}(2022)]%
        {wang2022frnet}
\bibfield{author}{\bibinfo{person}{Fangye Wang}, \bibinfo{person}{Yingxu Wang}, \bibinfo{person}{Dongsheng Li}, \bibinfo{person}{Hansu Gu}, \bibinfo{person}{Tun Lu}, \bibinfo{person}{Peng Zhang}, {and} \bibinfo{person}{Ning Gu}.} \bibinfo{year}{2022}\natexlab{}.
\newblock \showarticletitle{Enhancing CTR prediction with context-aware feature representation learning}. In \bibinfo{booktitle}{\emph{Proceedings of the 45th International ACM SIGIR Conference on Research and Development in Information Retrieval}}. \bibinfo{pages}{343--352}.
\newblock


\bibitem[Wang et~al\mbox{.}(2017)]%
        {wang2017deep}
\bibfield{author}{\bibinfo{person}{Ruoxi Wang}, \bibinfo{person}{Bin Fu}, \bibinfo{person}{Gang Fu}, {and} \bibinfo{person}{Mingliang Wang}.} \bibinfo{year}{2017}\natexlab{}.
\newblock \showarticletitle{Deep \& cross network for ad click predictions}.
\newblock In \bibinfo{booktitle}{\emph{Proceedings of the ADKDD'17}}. \bibinfo{pages}{1--7}.
\newblock


\bibitem[Wang et~al\mbox{.}(2021b)]%
        {wang2021dcn}
\bibfield{author}{\bibinfo{person}{Ruoxi Wang}, \bibinfo{person}{Rakesh Shivanna}, \bibinfo{person}{Derek Cheng}, \bibinfo{person}{Sagar Jain}, \bibinfo{person}{Dong Lin}, \bibinfo{person}{Lichan Hong}, {and} \bibinfo{person}{Ed Chi}.} \bibinfo{year}{2021}\natexlab{b}.
\newblock \showarticletitle{DCN v2: Improved deep \& cross network and practical lessons for web-scale learning to rank systems}. In \bibinfo{booktitle}{\emph{Proceedings of the web conference 2021}}. \bibinfo{pages}{1785--1797}.
\newblock


\bibitem[Wang et~al\mbox{.}(2021a)]%
        {wang2021masknet}
\bibfield{author}{\bibinfo{person}{Zhiqiang Wang}, \bibinfo{person}{Qingyun She}, {and} \bibinfo{person}{Junlin Zhang}.} \bibinfo{year}{2021}\natexlab{a}.
\newblock \showarticletitle{Masknet: Introducing feature-wise multiplication to CTR ranking models by instance-guided mask}.
\newblock \bibinfo{journal}{\emph{arXiv preprint arXiv:2102.07619}} (\bibinfo{year}{2021}).
\newblock


\bibitem[Yan et~al\mbox{.}(2022)]%
        {yan2022apg}
\bibfield{author}{\bibinfo{person}{Bencheng Yan}, \bibinfo{person}{Pengjie Wang}, \bibinfo{person}{Kai Zhang}, \bibinfo{person}{Feng Li}, \bibinfo{person}{Hongbo Deng}, \bibinfo{person}{Jian Xu}, {and} \bibinfo{person}{Bo Zheng}.} \bibinfo{year}{2022}\natexlab{}.
\newblock \showarticletitle{Apg: Adaptive parameter generation network for click-through rate prediction}.
\newblock \bibinfo{journal}{\emph{Advances in Neural Information Processing Systems}}  \bibinfo{volume}{35} (\bibinfo{year}{2022}), \bibinfo{pages}{24740--24752}.
\newblock


\bibitem[Yan et~al\mbox{.}(2014)]%
        {yan2014coupled}
\bibfield{author}{\bibinfo{person}{Ling Yan}, \bibinfo{person}{Wu-Jun Li}, \bibinfo{person}{Gui-Rong Xue}, {and} \bibinfo{person}{Dingyi Han}.} \bibinfo{year}{2014}\natexlab{}.
\newblock \showarticletitle{Coupled group lasso for web-scale ctr prediction in display advertising}. In \bibinfo{booktitle}{\emph{International conference on machine learning}}. PMLR, \bibinfo{pages}{802--810}.
\newblock


\bibitem[Yang et~al\mbox{.}(2022)]%
        {yang2022adasparse}
\bibfield{author}{\bibinfo{person}{Xuanhua Yang}, \bibinfo{person}{Xiaoyu Peng}, \bibinfo{person}{Penghui Wei}, \bibinfo{person}{Shaoguo Liu}, \bibinfo{person}{Liang Wang}, {and} \bibinfo{person}{Bo Zheng}.} \bibinfo{year}{2022}\natexlab{}.
\newblock \showarticletitle{Adasparse: Learning adaptively sparse structures for multi-domain click-through rate prediction}. In \bibinfo{booktitle}{\emph{Proceedings of the 31st ACM International Conference on Information \& Knowledge Management}}. \bibinfo{pages}{4635--4639}.
\newblock


\bibitem[Zhang et~al\mbox{.}(2022)]%
        {zhang2022dhen}
\bibfield{author}{\bibinfo{person}{Buyun Zhang}, \bibinfo{person}{Liang Luo}, \bibinfo{person}{Xi Liu}, \bibinfo{person}{Jay Li}, \bibinfo{person}{Zeliang Chen}, \bibinfo{person}{Weilin Zhang}, \bibinfo{person}{Xiaohan Wei}, \bibinfo{person}{Yuchen Hao}, \bibinfo{person}{Michael Tsang}, \bibinfo{person}{Wenjun Wang}, {et~al\mbox{.}}} \bibinfo{year}{2022}\natexlab{}.
\newblock \showarticletitle{DHEN: A deep and hierarchical ensemble network for large-scale click-through rate prediction}. In \bibinfo{booktitle}{\emph{Proceedings of the DLP-KDD}}.
\newblock


\bibitem[Zhang et~al\mbox{.}(2019)]%
        {zhang2019deep}
\bibfield{author}{\bibinfo{person}{Shuai Zhang}, \bibinfo{person}{Lina Yao}, \bibinfo{person}{Aixin Sun}, {and} \bibinfo{person}{Yi Tay}.} \bibinfo{year}{2019}\natexlab{}.
\newblock \showarticletitle{Deep learning based recommender system: A survey and new perspectives}.
\newblock \bibinfo{journal}{\emph{ACM computing surveys (CSUR)}} \bibinfo{volume}{52}, \bibinfo{number}{1} (\bibinfo{year}{2019}), \bibinfo{pages}{1--38}.
\newblock


\bibitem[Zhang et~al\mbox{.}(2024)]%
        {zhang2024multi}
\bibfield{author}{\bibinfo{person}{Yiqian Zhang}, \bibinfo{person}{Yinfu Feng}, \bibinfo{person}{Wen-Ji Zhou}, \bibinfo{person}{Yunan Ye}, \bibinfo{person}{Min Tan}, \bibinfo{person}{Rong Xiao}, \bibinfo{person}{Haihong Tang}, \bibinfo{person}{Jiajun Ding}, {and} \bibinfo{person}{Jun Yu}.} \bibinfo{year}{2024}\natexlab{}.
\newblock \showarticletitle{Multi-Domain Deep Learning from a Multi-View Perspective for Cross-Border E-commerce Search}. In \bibinfo{booktitle}{\emph{Proceedings of the AAAI Conference on Artificial Intelligence}}, Vol.~\bibinfo{volume}{38}. \bibinfo{pages}{9387--9395}.
\newblock


\bibitem[Zhou et~al\mbox{.}(2019)]%
        {zhou2019deep}
\bibfield{author}{\bibinfo{person}{Guorui Zhou}, \bibinfo{person}{Na Mou}, \bibinfo{person}{Ying Fan}, \bibinfo{person}{Qi Pi}, \bibinfo{person}{Weijie Bian}, \bibinfo{person}{Chang Zhou}, \bibinfo{person}{Xiaoqiang Zhu}, {and} \bibinfo{person}{Kun Gai}.} \bibinfo{year}{2019}\natexlab{}.
\newblock \showarticletitle{Deep interest evolution network for click-through rate prediction}. In \bibinfo{booktitle}{\emph{Proceedings of the AAAI conference on artificial intelligence}}, Vol.~\bibinfo{volume}{33}. \bibinfo{pages}{5941--5948}.
\newblock


\bibitem[Zhou et~al\mbox{.}(2018)]%
        {zhou2018deep}
\bibfield{author}{\bibinfo{person}{Guorui Zhou}, \bibinfo{person}{Xiaoqiang Zhu}, \bibinfo{person}{Chenru Song}, \bibinfo{person}{Ying Fan}, \bibinfo{person}{Han Zhu}, \bibinfo{person}{Xiao Ma}, \bibinfo{person}{Yanghui Yan}, \bibinfo{person}{Junqi Jin}, \bibinfo{person}{Han Li}, {and} \bibinfo{person}{Kun Gai}.} \bibinfo{year}{2018}\natexlab{}.
\newblock \showarticletitle{Deep interest network for click-through rate prediction}. In \bibinfo{booktitle}{\emph{Proceedings of the 24th ACM SIGKDD international conference on knowledge discovery \& data mining}}. \bibinfo{pages}{1059--1068}.
\newblock


\bibitem[Zhu et~al\mbox{.}(2018)]%
        {zhu2018learning}
\bibfield{author}{\bibinfo{person}{Han Zhu}, \bibinfo{person}{Xiang Li}, \bibinfo{person}{Pengye Zhang}, \bibinfo{person}{Guozheng Li}, \bibinfo{person}{Jie He}, \bibinfo{person}{Han Li}, {and} \bibinfo{person}{Kun Gai}.} \bibinfo{year}{2018}\natexlab{}.
\newblock \showarticletitle{Learning tree-based deep model for recommender systems}. In \bibinfo{booktitle}{\emph{Proceedings of the 24th ACM SIGKDD international conference on knowledge discovery \& data mining}}. \bibinfo{pages}{1079--1088}.
\newblock


\end{thebibliography}



\end{document}